\documentclass[preprint,showpacs,preprintnumbers,amsmath,amssymb,showkeys]{revtex4}

\usepackage[unicode,bookmarks,bookmarksopen,bookmarksopenlevel=0,colorlinks,%
pageanchor=1,breaklinks=1,linkcolor=blue,citecolor=blue,urlcolor=red,hypertex]{hyperref}

\usepackage{graphicx}
\usepackage{dcolumn}
\usepackage{bm}

\usepackage{float}
\newcommand{\Supp}{\operatorname{supp}}

\begin{document}

\preprint{APS/123-QED}


\title{Mathematical Model of a pH-gradient Creation at~Isoelectrofocusing. \\ Part I. Approximation of Weak Solution}

\author{L.\,V.~Sakharova}
\email{l_sakharova@mail.ru}
\affiliation{%
Institute of Water Transport \\ Rostov-on-Don, Russia
}%

\author{E.\,V.~Shiryaeva}
\email{shir@math.sfedu.ru}

\author{M.\,Yu.~Zhukov}%
 \email{myuzhukov@gmail.com}
\affiliation{%
Southern Federal University\\ Rostov-on-Don, Russia
}%

\date{\today}

\begin{abstract}
The mathematical model describing the stationary natural $\textrm{pH}$-gradient arising under the action of an electric field in an aqueous solution of ampholytes (amino acids)  is constructed and investigated. The model is a part of a more general model of the isoelectrofocusing process. Investigation is based on the approximation of a weak solution by the piecewise continuous non-smooth functions. The method can be used for solving classes of problems for ODEs with a small parameter at higher derivatives and the turning points.

\end{abstract}


\maketitle

\section{Introduction}\label{ZhS-01}

This paper starts a series of papers on the mathematical modeling of the isoelectrofocusing (IEF). These papers are linked by a common theme: the study of the
natural pH-gradient creation in aqueous solution of an amphoteric substances. We expect to investigate the following problems. 1. Onset of a piecewise constant pH-gradients
at large  values of voltage or electric current density, so called anomalous regimes. 2. Numerical study of the stationary IEF problem on the $\textrm{pH}$-gradient creation.
3. Numerical and analytical study of the non-stationary IEF problem on the $\textrm{pH}$-gradient creation. 4. The general mathematical IEF model and the specificities its construction. Each paper contains all the necessary information about the problem being solved and can be read independently.

Isoelectrofocusing (IEF) is a method of fractionation of multicomponent mixtures (proteins, peptides, amino acids) into individual components with the help of the electric field in a medium with non-uniform $\textrm{pH}$ distribution. The heart of the IEF method is an amphoteric properties of substances. Other words,
amino acid, proteins, and peptides have both acid and the base properties. At $\textrm{pH}=\textrm{pI}$, where $\textrm{pI}$ is so-called isoelectric point, the   electrophoretic mobility of substance equals zero. Isoelectric point $\textrm{pI}$ is the individual characteristic of amphoteric substance. In particular, almost all amino acids and peptides have various isoelectric points. This allows to identify them on values $\textrm{pI}$. In the presence of $\textrm {pH}$-gradient in electrophoretic chamber, components of the mixture move under the action of the external electric field until their electrophoretic mobility is equal to zero. As a result the spatial distribution of individual components exists on their values of $\textrm{pI}$.

The IEF method, along with the chromatography, the isotachophoresis, the zone electrophoresis, is  one of the most demanded methods of mixture fractionation in biology, chemistry, medicine. It is enough to tell that this method was widely used for Human Genome Project. Resolution of the IEF method, that is possibility of identification of the large quantity of the mixture component, depends on completeness of the information about $\textrm{pH}$-gradient.

There are various ways of creation $\textrm{pH}$-gradient in solution: creation of the artificial gradients with the help of the  special, so-called, buffer solution; creation of the immobilized gradients with the help of the organization of rigid chemical structures; creation of the natural gradients arising in solution as a result of a mixture components self-organization (for more details see \cite{ZhukovBabskiyYudovich,BabZhukYudE,MosherSavilleThorman,Righetti83,Righetti90,ZhRStoy2001}).

The natural $\textrm{pH}$-gradients are the most attractive from the point of view of simplicity using. These gradients were discovered by \cite{Vesterberg1966,Vesterberg1976,Haglund1971,Svensson1961} the first time theoretically and then experimentally. Long time it was assumed that the natural $\textrm{pH}$-gradients, more exact spatial distribution  of $\textrm {pH}$, are linear or, at least, are close to the linear. Distribution of substances in solution is close to Gaussian distribution. However, in 2004--2006 in \cite{Thormann2004,Thormann2006} with  the help of numerical integration of the non-stationary problem was revealed that the natural $\textrm {pH}$-gradients at large intensity of the external electric field in the stationary mode have step function profile and the distribution of concentration  closely to rectangular profiles. These results were partially confirmed by experiments. The theoretical explanation of the observed phenomenon was presented in \cite{Averkov,SakhVladZhuk}, where the rough asymptotic formulas describing step function $\textrm{pH}$ were obtained. Further, more exact relations were given, in particular, in \cite{SakhSKNC,SakhOrel}.

From the mathematical point of view, the modeling of stationary natural $\textrm {pH}$-gradients problem  is reduced to the solution of the ODE's equations for distribution of concentration, some algebraic constrain and integral conditions. At large intensity of the electric field (or large density of an electric current) the system of the equations is stiff: ODE's have the small parameter at the highest  derivatives and have the turning points. Numerical integration of this problem becomes complicated also that solutions for separate concentration are focused in some regions of the integration interval and quickly exponential decrease out of these regions.

In this paper the approximate method based on approximation of the weak solution by piecewise continuous functions is developed. The various approximations of solution are presented and the error estimates are given. Such method can be used for the solution of classes of problem with small parameter at the highest derivatives and a large number of turning points.

The paper is organized as follows. In Sec.~\ref{ZhS-02} the general equations of electrophoresis are described.
In Sec.~\ref{ZhS-03} the basic stationary equations governing the IEF process and $\textrm{pH}$-gradient are included.
In Sec.~\ref{ZhS-04} the weak formulation of the origin problem is given. In Secs.~\ref{ZhS-05}--\ref{ZhS-10} the piecewise-smooth approximation of weak solutions, the choice of the approximating functions, the algorithm for the approximation of weak solutions, and examples of approximation are presented. In Sec.~\ref{ZhS-11} other way of the approximating functions selecting is given. In Sec.~\ref{ZhS-12} the weak solutions approximation at moderate parameter is demonstrated.
Appendix~\ref{App:ZhS-1} contains the method of the integral asymptotic evaluation.
Appendix~\ref{App:ZhS-5} contains the generalized solution of the problem for limiting case.

\section{General Equations}\label{ZhS-02}

The general non-stationary equations (in dimensionless variables) describing process of creation natural $\rm{pH}$- gradient in multicomponent chemically active media have the following
form (see, \cite{ZhukovBabskiyYudovich,BabZhukYudE,MosherSavilleThorman,Zhukov2005,SakhVladZhuk}):
\begin{equation}\label{ZhS-eq1}
\partial_t a_{k}+\operatorname{div} \boldsymbol{i}_k=0,\quad \boldsymbol{i}_k = -\varepsilon\mu_{k} \nabla a_k+\mu_{k} e_{k}(\psi)a_k \boldsymbol{E},
\quad k=1,\dots,n,
\end{equation}
\begin{equation}\label{ZhS-eq2}
   \sum_{k=1}^{n}e_{k}(\psi)a_{k}=0,
\end{equation}
\begin{equation}\label{ZhS-eq3}
\boldsymbol{j}=\sum_{k=1}^{n}
\left(
-\varepsilon \mu_k\nabla(e_{k}(\psi)a_k)+\mu_{k}\sigma_k(\psi)a_k \boldsymbol{E}
\right), \quad
 \operatorname{div}\boldsymbol{j}=0,
\end{equation}
where $a_k$, $\boldsymbol{i}_k$ are the analytical concentration and the flux density of the components, $\boldsymbol{E}$ is the intensity of external electric field, $\boldsymbol{j}$ is the density of the electric current, $\psi$ is the acidity function of the mixture, $\mu_{k} e_{k}(\psi)$, $\mu_{k}\sigma_{k}(\psi)$, $\mu_{k}>0$, $\varepsilon \mu_k$ are the electrophoretic mobility, partial conductivity, characteristic mobility and diffusion coefficient of the components.

Used in chemistry function $\textrm{pH}$ is connected with concentration of hydrogen ions and acidity function $\psi$ by relations:
\begin{equation*}
\textrm{pH}=-\lg [\textrm{H}^+], \quad
[\textrm{H}^+]=K_w e^{\psi}, \quad \textrm{pH}=-\lg K_w-\psi\lg e,
\end{equation*}
where $[\textrm{H}^+]$ is the concentration of hydrogen ions ($\text{\rm {mol}/\rm{l}}$), $K_w=10^{-7}\,\text{\rm{mol}/\rm{l}}$ is the autodissociation constant of water.

The equations (\ref{ZhS-eq1}) are the usual diffusion equations with transport under action of the electric field. The algebraic equation (\ref{ZhS-eq2}) is the  electroneutrality condition. The equation (\ref{ZhS-eq3}) is the general Ohm law.

To close the equations system (\ref{ZhS-eq1})--(\ref{ZhS-eq3}) we define the dependence of electrophoretic mobility  and partial conductivity on $\psi$, i.e. functions $e_k=e_k(\psi)$, $\sigma_k =\sigma_k(\psi)$.

In case of the mixture of amphoteric substances the dissociation reactions have the following form (see, for example, \cite{ZhukovBabskiyYudovich,BabZhukYudE,SakhVladZhuk}):
\begin{equation*}
   \textrm{H}^+ \textrm{R}
  \overset{B_i}{\rightleftharpoons}
  \textrm{R}_i^0 + \textrm{H}^+,
  \quad
  \textrm{R}^{0}_i
  \overset{A_i}{\rightleftharpoons}
     \textrm{R}^{-}_i + \textrm{H}^+.
\end{equation*}
Here, $\textrm {R}_i^0$ is zwitterion (`neutral' ion), $A_i$ and $B_i$ are the dissociation constants for acid ($\textrm {R}_i^-$) and base ($\textrm {H}^+\textrm{R}_i$) groups, $\textrm{H}^+$ is the hydrogen ion.

For example, for amino acid $\textrm{NH}_3^+\textrm {R}\textrm{COO}^-$,
where $\textrm{NH}_3^+$ is the amino group, $\textrm {R}$ is amino acid residue, $\textrm {COO}^-$ is the carboxyl group, we have:
$\textrm{H}^+\textrm{R}\equiv \textrm{NH}_3^+\textrm{R}\textrm{COOH}$,
$\textrm{R}^{-}_i\equiv\textrm{NH}_2\textrm{R}\textrm{COO}^-$,
$\textrm{R}^{0}_i\equiv \textrm{NH}_3^+\textrm{R}\textrm{COO}^-$.

The specified reactions proceed almost instantly and balance conditions of this reactions allow to determine dependence of electrophoretic mobility and partial conductivity on acidity function $\psi$ \cite{ZhukovBabskiyYudovich,BabZhukYudE,Zhukov2005}:
\begin{equation}\label{ZhS-eq4}
e_{i}(\psi)=\frac{[\textrm{H}^+\textrm{R}_i]-[\textrm{R}^{-}_i]}{a_i}, \quad
\sigma_{i}(\psi)=\frac{[\textrm{H}^+\textrm{R}_i]+[\textrm{R}^{-}_i]}{a_i},\quad
a_i=[\textrm{H}^+\textrm{R}_i] + [\textrm{R}^0_i] +[\textrm{R}^{-}_i],
\end{equation}
\begin{equation*}
   e_{i}(\psi)= \frac{\sinh(\psi-\psi_{i})}{\cosh(\psi-\psi_{i})+\delta_{i}}, \quad
   \sigma_{i}(\psi)=\frac{\cosh(\psi-\psi_{i})}{\cosh(\psi-\psi_{i})+\delta_{i}}, \quad
   \psi_i=\frac12 \ln\frac{A_i B_i}{K_w^2}, \quad \delta_i=\frac12 \sqrt{\frac{B_i}{A_i}},
\end{equation*}
where $\delta_i>0$ is the dimensionless parameter, $\psi_i$ is the isoelectric point (electrophoretic mobility $\mu_i e_i$ is equal to zero at $\psi=\psi_i$, \emph{i.e.} $\mu_i e_i(\psi_i)=0$).

Note the important role of the electroneutrality condition for the description of transport process  in chemically active media. The algebraic equation (\ref {ZhS-eq2}) defines the function  $\psi$. Actually, it is the instant regulator of process. Permutations of the component concentrations $a_k$ lead to change of acidity function $\psi$. In turn, the kinetic coefficients of $e_{k}(\psi)$, $\sigma_{k}(\psi)$ influence on transport of the component $a_k$.

Finally, we specify connection between dimensional and dimensionless variables:
\begin {equation*}
  \widetilde{x} = xL_*, \quad
  \widetilde{t} = tt_*, \quad
  \widetilde{a} _k = a_k C_*, \quad
  \widetilde{E} = E E_*, \quad
  \widetilde{j} = j F_* C_* E_* \mu_*,
\end {equation*}
\begin {equation*}
  \varepsilon = \frac {R_*T_*}{F_* E_* L_*}, \quad
   {t_*} = \frac {L_*} {E_*\mu_*}.
\end {equation*}
Here, $L_*$, $t_*$, $E_*$, $C_*$ are the characteristic length, time, intensity of the electric field and analytical concentration; $\mu_*$ is the characteristic mobility; $F_*\approx 96485.34\,\, \rm{C}\cdot \rm{mol}^{-1}$ is the Faraday's number, $R_*\approx 8.314462\,\, \rm{J}\cdot\rm{mol}^{-1}\cdot\rm{K}^{-1}$ is the universal gas constant, $T_*$ is the absolute temperature of the mixture.

In practice  of IEF the voltage $ E_* L_*$ changes usually from $1\,\,\rm{kV}$ to $10\,\,\rm{kV}$ and temperature is $T_*\approx 293\,\,\rm{K}$. In this case parameter $\varepsilon$ changes from $2.5\cdot 10^{-5}$ to $2.5\cdot 10^{-6}$.

\section{Stationary problem}\label{ZhS-03}

We formulate the problem for definition of the stationary natural  $\textrm {pH}$-gradient in the one-dimensional case. The one-dimensional case is the most demanded because usually for IEF the cylindrical electrophoretic chamber is used. In other cases, for IEF the flat thin plates are used for which the characteristic size in the direction of an electric field action much more then other plate sizes \cite {Righetti83,Righetti90}. Information about stationary $\textrm {pH}$-gradient is most important for interpretation of an experimental results. Of course, for obtaining the stationary solution of the equations (\ref{ZhS-eq1})--(\ref{ZhS-eq4}) the numerical integration of the non-stationary problem can be used (see, \cite{Thormann2004,Thormann2006}). Such method is good because it allows to trace dynamics of process. However, for large numbers of mixture components the numerical integration of the non-stationary problem requires a long times. It is obvious that instead of use the numerical integration of the non-stationary problem it is rather directly to solve the stationary problem.

We require the impermeability condition on the boundary of the electrophoretic chamber ($0 \leqslant x \leqslant  L$):
\begin{equation}
   i_{k}\bigr|_{x=0,L}=0,\quad k=1,\dots,n.
\label{ZhS-eq5}
\end{equation}

For the one-dimensional case the solution of the  electric current continuity  equation (\ref{ZhS-eq3}) is $j=j(t)$. For a stationary problem it is naturally to consider
\begin {equation}
   j(t)=j_0,
   \label{ZhS-eq6}
\end {equation}
where $j_0$ is the constant electric current density.

Strictly speaking, in dimensionless variables length of the electrophoretic chamber is $L=1$ and the electric current density is $j_0=1$. However, for interpretation of results using $L$ and $j_0$ is more convenient.

The problem (\ref{ZhS-eq1})--(\ref {ZhS-eq5}) for definition of the functions $a_k(x) $, $k=1,\dots,n$, $\psi(x)$  has the following form:
\begin{equation}\label{ZhS-eq7}
  \frac{1}{\lambda}\frac{da_k}{dx} =
  \frac{a_k\theta_k(\psi)\sum\limits_{i=1}^n a_i\theta'_i(\psi)}
  {\sigma\sum\limits_{i=1}^n  a_i \left( \theta^2_i(\psi) + \theta'_i(\psi)\right)}, \quad  k=1,...,n,
  \quad  0 \leqslant x \leqslant L,\quad \lambda=\frac{j_0}{\varepsilon},
\end{equation}
\begin{equation}\label{ZhS-eq8}
  \sum\limits_{k=1}^n a_k \theta_k(\psi) = 0,
\end{equation}
\begin{equation}\label{ZhS-eq9}
   \int\limits_0^L a_k(x)\,dx = M_k,
\end{equation}
\begin{equation}\label{ZhS-eq10}
 \sigma =\sum_{i=1}^n \mu_i a_i \theta'_i(\psi), \quad \theta_i(\psi)=\frac{\varphi'_i(\psi)}{\varphi_i(\psi)},
 \quad \varphi_i(\psi)=\cosh(\psi-\psi_i)+\delta_i,
\end{equation}
where $M_k$ is the quantity  of $a_k$  on the interval $[0,L]$.

The additional conditions (\ref{ZhS-eq9}) are implication of mass conserve law. We add these conditions because conditions
(\ref{ZhS-eq5}) are not enough to solve the stationary problem.

The detailed description of transition from the equations (\ref{ZhS-eq1})--(\ref{ZhS-eq5}) to the equations (\ref{ZhS-eq7})--(\ref{ZhS-eq10}) contains in
\cite{SakhVladZhuk,SakhSKNC,SakhOrel}. Here, we only specify that for such transition it is enough to present the equation (\ref{ZhS-eq3}) in the form
$j =\sigma (E-\varepsilon\psi')$ and then exclude $(E-\varepsilon\psi')$ from the equations.

The system (\ref{ZhS-eq7})--(\ref{ZhS-eq10}) has integral which one can get by the summation of all equations (\ref{ZhS-eq7}) and taking into account (\ref{ZhS-eq8}):
\begin{equation}\label{ZhS-eq11}
 \sum_{i=1}^n a_i =a_0\equiv L^{-1}\sum\limits_{i=1}^n M_i,
\end{equation}
where the constant $a_0$ is defined by (\ref{ZhS-eq9}).

We note that $\psi(x)$ is a monotone decreasing function. This property is  easy to get by differentiating the electroneutrality equation (\ref{ZhS-eq8}) at the assumption of a sufficient smoothness:
\begin{equation}\label{ZhS-eq12}
   \frac{d\psi}{dx}=-\frac{\lambda\sum\limits_{i=1}^n a_i\theta_i^2(\psi)}
  {\sigma\sum\limits_{i=1}^n  a_i \left(\theta^2_i(\psi) + \theta'_i(\psi)\right)}<0.
\end{equation}
The negativity of the derivative follows from the relations (\ref{ZhS-eq10}). In fact, it is easy to show that $(\theta^2_i(\psi) + \theta'_i(\psi)>0$ and the functions $a_k(x)$ not equal to zero simultaneously.

As already mentioned, the solution of (\ref{ZhS-eq7})--(\ref{ZhS-eq11}) for large values of the parameter $\lambda$ involves difficulties due to the presence of a small parameter at highest derivatives and the turning points at $\psi=\psi_i$. Preliminary numerical analysis shows that for large values of $\lambda$ the concentrations are localized in some segment of the interval $[0,L]$ (each in the own segment) and exponentially decreasing outside these segments. It means that the using for numerical integration, for example, the shooting method (the transform the boundary problem to the  Cauchy problem) in combination with the Newton's is complicated. In fact, the initial conditions at one of the ends of the segment are the order of $O(e^{-\lambda})$ and for their determination a very detailed initial approximation is required (see \cite{SakhVladZhuk,SakhSKNC,SakhOrel}). However, for example, in \cite{SakhVladZhuk} it is shown that the asymptotic solutions tend to some generalized functions: the profile of the concentrations $a_k(x)$ has almost rectangular shape. Such behavior of the solutions, as will be shown below, allows to construct a continuous piecewise-smooth approximation of solutions, refusing from the function smoothness and going to the weak formulation of the problem.

The main goal of this paper is the construction of the  piecewise continuous approximation of a weak solution of the problem (\ref{ZhS-eq7})--(\ref{ZhS-eq11}) for given parameters $\mu_k$, $\delta_k$, $M_k$, $k=1,\dots,n$, which have order $O(1)$, and the large parameter $\lambda \gg 1$.

\section{The weak formulation of the problem (\ref{ZhS-eq7})--(\ref{ZhS-eq11})}\label{ZhS-04}

As usual, we call the weak solution of the problem (\ref{ZhS-eq7})--(\ref{ZhS-eq11})  the functions $a_k(x)$, $k=1,\dots,n$, $\psi(x)$ satisfying the equations:
\begin{equation}
I_k\equiv\int\limits_0^L
\left(a_k\frac{dV_k}{dx} +
\frac{\lambda a_k \theta_k(\psi)\sum\limits_{i=1}^n a_i\theta'_i(\psi)}{\sigma\sum\limits_{i=1}^n  a_i \left(\theta^2_i(\psi) + \theta'_i(\psi)\right)}
 V_k
\right)dx=0,
\label{ZhS-eq13}
\end{equation}
\begin{equation*}
V_k(0)=0, \quad V_k(L)=0,
\end{equation*}
\begin{equation}\label{ZhS-eq14}
  \sum\limits_{k=1}^n a_k \theta_k(\psi) = 0, \quad
\end{equation}
\begin{equation}\label{ZhS-eq15}
  \int\limits_0^L a_k(x)\,dx = M_k,
\end{equation}
\begin{equation}\label{ZhS-eq16}
  \sum_{i=1}^n a_i =a_0\equiv L^{-1}\sum\limits_{i=1}^n M_i,
\end{equation}
\begin{equation*}
 \sigma =\sum_{i=1}^n \mu_i a_i \theta'_i(\psi), \quad \theta_i(\psi)=\frac{\varphi'_i(\psi)}{\varphi_i(\psi)},
 \quad \varphi_i(\psi)=\cosh(\psi-\psi_i)+\delta_i.
\end{equation*}
Here, $V_k(x)$ are arbitrary sufficiently smooth functions satisfying the natural boundary conditions.

Note that the relations (\ref{ZhS-eq8}) and (\ref{ZhS-eq9}), \emph{i.e.} the electroneutrality condition and conditions of the mass conservation, remain the same. The relation (\ref{ZhS-eq16}), \emph{i.e.} the integral (\ref{ZhS-eq11}) of the system (\ref{ZhS-eq7}), is not implementation of the system (\ref{ZhS-eq13}). For the weak formulation of the problem the relation (\ref{ZhS-eq16}), in principle, can be discarded. The most reasonable, of course, to assume that the condition (\ref{ZhS-eq16}) holds, thus preserving some additional properties of the original problem (\ref{ZhS-eq7})--(\ref{ZhS-eq11}).

Naturally, in the case when the functions $a_k(x)$, $k=1,\dots,n$, $\psi(x)$ are sufficient smoothness the weak solution of (\ref{ZhS-eq13})--(\ref{ZhS-eq16}) will be the strong solution of the original problem (\ref{ZhS-eq7})--(\ref{ZhS-eq11}).

\section{Piecewise-smooth approximation of weak solutions}\label{ZhS-05}

\begin{figure}[H]
\centering
\includegraphics[scale=0.85]{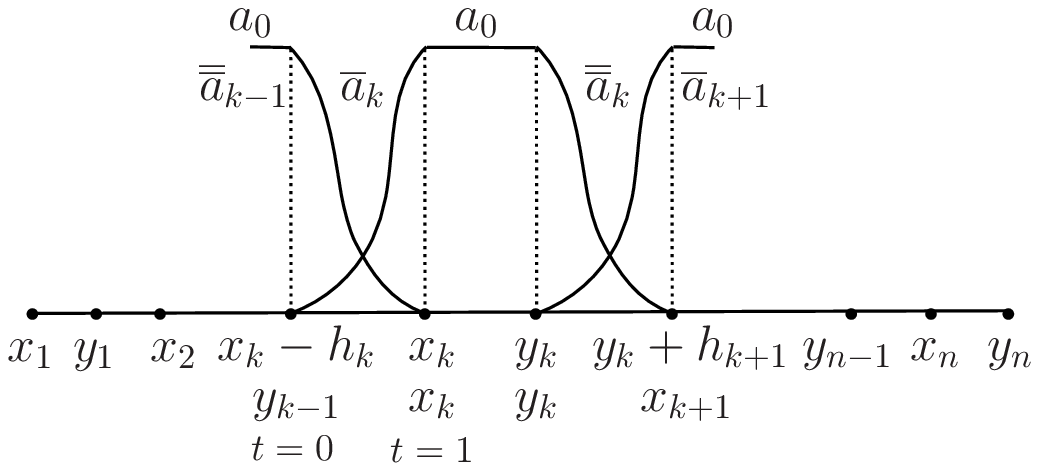}\\ 
\includegraphics[scale=0.85]{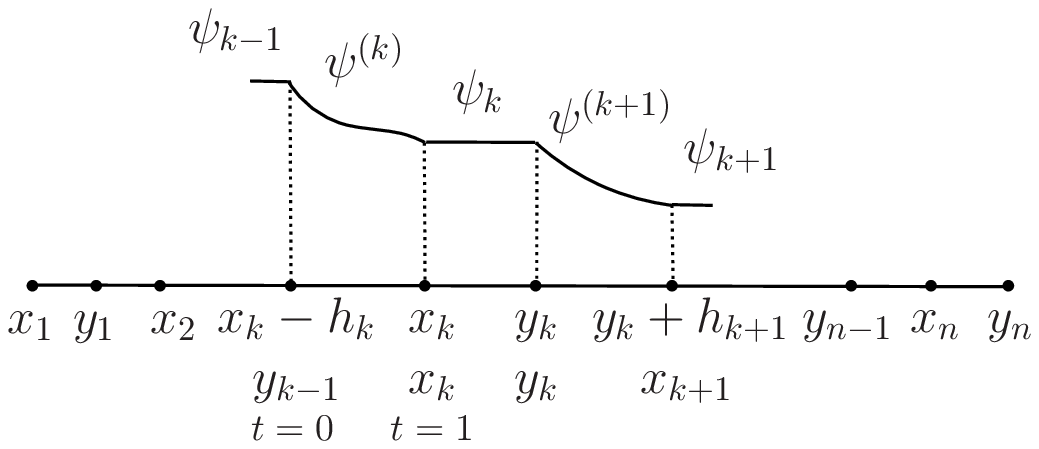}\\ 
\caption{Scheme of approximation}
\label{Ris1}
\end{figure}

We define the partition of interval $[0,L]$ by the set of points (see Fig.~\ref{Ris1})
\begin{equation*}
0=x_1 < y_1 < x_2 < y_2 <\dots <y_{k-1}< x_k < y_k < x_{k+1}<\dots < y_{n-1} < x_n < y_n=L.
\end{equation*}
The method of selection of the points $x_k$, $y_k$ is specified in section \ref{ZhS-09}.

We choose the functions $a_k$ satisfying the following properties:
\begin{equation*}
\Supp a_k=[x_k-h_k, y_k+h_{k+1}] \subset [0,L], \quad k=1,\dots,n,
\end{equation*}
\begin{equation*}
\quad h_1=0, \quad h_k=x_k-y_{k-1}, \quad k=2,\dots,n, \quad h_{n+1}=0.
\end{equation*}
It is obvious that
\begin{equation}\label{ZhS-eq17}
\Supp a_{k-1} \cap \Supp a_k =[x_k-h_k, x_k],
\end{equation}
\begin{equation*}
a_i=0, \quad (i\neq k-1,k, \quad x\in [x_k-h_k, y_k+h_{k+1}].
\end{equation*}

To solve problem (\ref{ZhS-eq13})--(\ref{ZhS-eq16}) we use approximation (see Fig.~\ref{Ris1})
\begin{equation} \label{ZhS-eq18}
 a_k(x)= \left\{
 \begin{array}{ll}
 0,                            &  x\leqslant x_k-h_k, \\
 \overline{a}_k(x),            &  x_k-h_k \leqslant  x\leqslant x_k, \\
 a_0,                          &  x_k \leqslant x \leqslant y_k, \\
 \overline{\overline{a}}_k(x), &  y_k \leqslant x \leqslant y_k+h_{k+1}, \\
 0,                            &  y_k+h_{k+1} \leqslant x,
  \end{array}
        \right.
        \quad k=1,\dots,n,
\end{equation}
\begin{equation} \label{ZhS-eq19}
 \psi(x)= \left\{
 \begin{array}{ll}
\psi^{(k)}(x),             &  x_k-h_k \leqslant x \leqslant x_k, \\
 \psi_k,                        &  x_k \leqslant x \leqslant  y_k,\\
 \psi^{(k+1)}(x), &  y_k \leqslant x  \leqslant y_k+h_{k+1}.
  \end{array}
          \right.
          \quad k=2,\dots,n.
\end{equation}
 Here, $\overline{a}_k(x)$, $\overline{\overline{a}}_k(x)$, $\psi^{(k)}(x)$, $\psi^{(k+1)}(x)$ are the enough smooth function (at appropriate intervals) satisfying to continuity conditions:
 \begin{equation}\label{ZhS-eq20}
  \overline{a}_k(x_k-h_k)=0, \quad  \overline{a}_k(x_k)=a_0, \quad
  \overline{\overline{a}}_k(y_k)=a_0, \quad  \overline{\overline{a}}_k(y_k+h_{k+1})=0,
 \end{equation}
 \begin{equation}\label{ZhS-eq21}
 \psi^{(k)}(x_k-h_k)=\psi_{k-1}, \quad  \psi^{(k)}(x_k)=\psi_k,
\end{equation}
 \begin{equation*}
 \psi^{(k+1)}(y_k)=\psi_k, \quad  \psi^{(k+1)}(y_k+h_{k+1})=\psi_{k+1}.
\end{equation*}

\section{The reduction of the integrals $I_k$}\label{ZhS-06}

We introduce notations for integrand functions:
\begin{equation}
G_k(a;\psi)\equiv
a_k\frac{dV_k}{dx} +
\frac{\lambda a_k \theta_k(\psi)\sum\limits_{i=1}^n a_i\theta'_i(\psi)}
{\sigma(a;\psi)\sum\limits_{i=1}^n a_i
\left(
\theta^2_i(\psi) + \theta'_i(\psi)
\right)}
V_k,
\label{ZhS-eq22}
\end{equation}
\begin{equation}
F_k(a;\psi)\equiv
-\frac{da_k}{dx} +
\frac{\lambda a_k \theta_k(\psi)\sum\limits_{i=1}^n a_i\theta'_i(\psi)}
{\sigma(a;\psi)\sum\limits_{i=1}^n a_i
\left(
\theta^2_i(\psi) + \theta'_i(\psi)
\right)},
\label{ZhS-eq23}
\end{equation}
\begin{equation}
\sigma(a;\psi)\equiv
\sum\limits_{i=1}^n \mu_i a_i\theta'_i(\psi, \quad a=(a_1,a_2,\dots,a_n).
\label{ZhS-eq24}
\end{equation}

The choice of $a_k(x)$, $\psi(x)$ in the form (\ref{ZhS-eq18}), (\ref{ZhS-eq19}) allows to write integrals (\ref{ZhS-eq13}) in the form:
\begin{equation*}
I_k =\int\limits_{x_k-h_k}^{x_k}
G_k(a;\psi^{(k)})\,dx +
\int\limits_{x_k}^{y_k} a_0\frac{dV_k}{dx} +
\int\limits_{y_k}^{y_k+h_{k+1}}
G_k(a;\psi^{(k+1)})\,dx.
\end{equation*}
Here, we take into account the relations $\theta_k(\psi_k)=0$.

The smoothness of the functions $\overline{a}_k$, $\overline{\overline{a}}_k$,
$\psi^{(k)}_k$, $\psi^{(k+1)}_k$ allows to use integration by parts. Taking into account (\ref{ZhS-eq20}) we omit all integrated term
and $I_k$ take the form:
\begin{equation}
I_k =\int\limits_{x_k-h_k}^{x_k}F_k(a;\psi^{(k)})V_k\,dx
+\int\limits_{y_k}^{y_k+h_{k+1}}F_k(a;\psi^{(k+1)})V_k\,dx.
\label{ZhS-eq25}
\end{equation}

\section{The choice of the approximating functions}\label{ZhS-07}

The functions $a_k(x)$, $k=1,\dots,n$, $\psi(x)$ are defined by the relations (\ref{ZhS-eq18}), (\ref{ZhS-eq19}) will be the solution of the problem
(\ref{ZhS-eq13})--(\ref{ZhS-eq16}) if $I_k \to 0$ at $\lambda \to \infty$.

Using the special selection of functions $\psi^{(k)}(x)$, $\overline{a}_k(x)$,  $\overline{\overline{a}}_k(x)$ we show that  $I_k \to 0$ at $\lambda \to \infty$.

We focus only on the first integral of (\ref{ZhS-eq25}), \emph{i.e.} the integral over the interval $[x_k-h_k,x_k]$. For the second integral all of the arguments remain valid.

It is convenient to change variables:
\begin{equation}\label{ZhS-eq26}
x=x_k-h+th_k, \quad 0\leqslant t \leqslant 1, \quad dx=h_k dt.
\end{equation}
Then the first integral (\ref{ZhS-eq25}) has the form
\begin{equation}\label{ZhS-eq27}
I_k^0=\int\limits_0^1 F_k(t)
V_k(x_k-h_k+th_k) h_k\,dt,
\end{equation}
where (see (\ref{ZhS-eq23}))
\begin{equation}\label{ZhS-eq28}
F_k(t)=F_k(a(t);\psi^{(k)}(t))=
-\frac{da_k}{dt}\frac{1}{h_k} +
\frac{\lambda a_k \theta_k(\psi^{(k)})\sum\limits_{i=k-1}^k\!\!a_i\theta'_i(\psi^{(k)})}
{\sigma(a(t);\psi^{(k)})\sum\limits_{i=k-1}^k\!\!a_i
\left(
\theta^2_i(\psi^{(k)}) + \theta'_i(\psi^{(k)})
\right)}.
\end{equation}

We omitted the `overline' symbol, \emph{i.e.} $\overline{a}_k=a_k$, $\overline{\overline{a}}_{k-1}=a_{k-1}$. For functions $a_m(x_k-h+th)$, $m=k-1,k$, $\psi^{(k)}(x_k-h+th)$ after substitution (\ref{ZhS-eq26}) we use previous notation
\begin{equation*}
a_m(t)=a_m(x_k-h_k+th_k), \quad \psi^{(k)}(t)=\psi^{(k)}(x_k-h_k+th_k).
\end{equation*}

Note, conditions (\ref{ZhS-eq17}) means that only the functions $a_{k-1}$ and $a_{k}$ are not equal zero on the interval $[x_k-h_k, x_k]$. We use this fact writing the formula (\ref{ZhS-eq28}).

Again, taking into account conditions (\ref{ZhS-eq17}) we get (\ref{ZhS-eq14}), (\ref{ZhS-eq16}) on the interval $[x_k-h_k, x_k]$ as:
\begin{equation}\label{ZhS-eq29}
a_{k-1}+a_k=a_0, \quad \theta_{k-1} a_{k-1}+\theta_k a_k=0.
\end{equation}

The linear system of equations (\ref{ZhS-eq29}) allows to easily determine the dependence of the $a_{k-1}$, $a_{k}$ on $\psi^{(k)}$:
\begin{equation}\label{ZhS-eq30}
a_{k-1}(t)=\frac{a_0 \theta_k(\psi^{(k)}(t))}{ \theta_k(\psi^{(k)}(t))- \theta_{k-1}(\psi^{(k)}(t))},\quad
a_{k}(t)=-\frac{a_0 \theta_{k-1}(\psi^{(k)}(t))}{ \theta_k(\psi^{(k)}(t))- \theta_{k-1}(\psi^{(k)}(t))}.
\end{equation}

Substitution (\ref{ZhS-eq29}) into (\ref{ZhS-eq28}) and substitution $F_k$ into (\ref{ZhS-eq27}) shows that the integral $I_k^0$ is a nonlinear functional $I_k^0=I_k^0[\psi^{(k)}]$. It means that to obtain the required result: $I_k^0[\psi^{(k)}]\to 0$ at $\lambda \to \infty$, it is enough to choose only function $\psi^{(k)}(t)$.

The function $\psi^{(k)}(t)$ must be a monotonically decreasing function satisfying to the conditions (\ref{ZhS-eq21}):
\begin{equation}\label{ZhS-eq31}
\psi^{(k)}(t)\bigr|_{t=0}=\psi_{k-1}, \quad \psi^{(k)}(t)\bigr|_{t=1}=\psi_{k}, \quad \psi'(t)<0.
\end{equation}
The requirement of monotonic decreasing functions $\psi^{(k)}(t)$ is dictated by the monotonicity condition of the respective function for the original problem (see (\ref{ZhS-eq12})).

Note that condition (\ref{ZhS-eq31}) automatic imply the  conditions corresponding to (\ref{ZhS-eq20}):
\begin{equation}\label{ZhS-eq32}
a_{k-1}(t)\bigr|_{t=0}=a_0, \quad
a_{k-1}(t)\bigr|_{t=1}=0, \quad
a_{k}(t)\bigr|_{t=0}=0, \quad
a_{k}(t)\bigr|_{t=1}=a_0.
\end{equation}

The natural constraints on the choice of the function $\psi^{(k)}(t)$ is imposed by the condition of the existence of integral $I_k^0$ and the integrals in (\ref{ZhS-eq16}).

Unfortunately, we cannot choose a function $\psi^{(k)}(t)$  so that the condition
$F_k(t)=0$ will be valid. Analysis shows that the requirement of $F_k(t)=0$
is equivalent to the equation (\ref{ZhS-eq12}). In this case the integrals in (\ref{ZhS-eq16}) have the singularities.


\section{Evaluation of integrals $I^0_k$}\label{ZhS-08}

We show that the appropriate choice of $\psi^{(k)}(t)$ allows to obtain the estimate $h_k=O(\lambda^{-1})$ and $I_k^0=O(\lambda^{-1})$ at $\lambda\to \infty$ .

Using infinite differentiability of functions $V_k$ and the Taylor series expansion in a neighborhood of some point $t=t_0$ (or for old variables $\overline{x}=x_k-h_k+t_0h_k$) for the integral (\ref{ZhS-eq27}) we have:
\begin{equation}\label{ZhS-eq33}
I_k^0=h_k V_k(\overline{x}) \int\limits_0^1 F_k(t)\,dt + h_k^2 \frac{d V_k(\overline{x})}{dx}\int\limits_0^1 F_k(t)(t-t_0)\, dt+\cdots.
\end{equation}

The rough estimate of the function $F_k(t)$ is $F_k(t)=O(\lambda)$. This means that the first term in (\ref{ZhS-eq33}) has the order $O(h_k\lambda)$ and is not small when $\lambda\to 0$ even if $h_k=O(\lambda^{-1})$.

To destroy the first term in (\ref{ZhS-eq33}) we require
\begin{equation}\label{ZhS-eq34}
\int\limits_0^1 F_k(t) dt=0.
\end{equation}
Then
\begin{equation}\label{ZhS-eq35}
I_k^0= h_k^2 \frac{d V_k(\overline{x})}{dx}\int\limits_0^1 F_k(t) (t-t_0) dt+...
\end{equation}

Using the requirements (\ref{ZhS-eq34}) and (\ref{ZhS-eq28}) we get
\begin{equation}\label{ZhS-eq36}
\frac{1}{h_k}\int\limits_0^1 \frac{da_k(t)}{dt}\,dt=\lambda\int\limits_0^1 \Phi_k(t)\,dt,
\end{equation}
where
\begin{equation}\label{ZhS-eq37}
\Phi_k(t)=\frac{a_k(t) \theta_k(\psi^{(k)}(t))\sum\limits_{i=k-1}^k\!\!a_i(t)\theta'_i(\psi^{(k)}(t))}
{\sigma(a(t);\psi^{(k)}(t))\sum\limits_{i=k-1}^k\!\!a_i(t)
\left(
\theta^2_i(\psi^{(k)}(t)) + \theta'_i(\psi^{(k)}(t))
\right)}.
\end{equation}
Finally, taking into account (\ref{ZhS-eq32}) we rewrite (\ref{ZhS-eq36}) as:
\begin{equation}\label{ZhS-eq38}
h_k=\frac{a_0}{\lambda\int_0^1 \Phi_k(t)\,dt}.
\end{equation}

Thus, the special choice of the monotonically decreasing function $\psi^{(k)}(x)$ satisfying to (\ref{ZhS-eq31}) implies the relations:
\begin{equation}\label{ZhS-eq39}
\int\limits_0^1 \Phi_k(t)\,dt=O(1), \quad h_k\int\limits_0^1 a_k(t)dt=O(1)
\end{equation}
and
\begin{equation}\label{ZhS-eq40}
h_k=O(\lambda^{-1}), \quad I_k^0=O(\lambda^{-1}).
\end{equation}

The last estimates  mean that the approximation (\ref{ZhS-eq18}), (\ref{ZhS-eq19}) is a weak solution of (\ref{ZhS-eq13})--(\ref{ZhS-eq16}) at $\lambda\to \infty$.

\section{The algorithm for the approximation of weak solutions}\label{ZhS-09}

Here, we present a simple algorithm for constructing an approximation (\ref{ZhS-eq18}), (\ref{ZhS-eq19}). We assume that the parameters $\psi_k$, $\delta_k$, $\mu_k$, $M_k$, $k=1,\dots,n$, $L$ are given, $a_0$ is defined by (\ref{ZhS-eq11}), and the parameter $\lambda$ is large enough.

1. On each, while unknown, interval $[x_k-h_k,x_k]$, $k=2,\dots,n$ we choose some monotonically decreasing function $\psi^{(k)}(t)$ satisfying to the conditions
(\ref{ZhS-eq31}). According to the formulae (\ref{ZhS-eq30}) we define the function $a_{k-1}(t)$, $a_k(t)$, $k=2,\dots,n$ on each interval $[x_k-h_k,x_k]$.
Using equation (\ref{ZhS-eq37}), (\ref{ZhS-eq38}) we calculate the lengths of segments $h_k$, $k=2,\dots,n$.

2. On each interval $[x_k-h_k,x_k]$ we calculate the $m_{k-1}^{(k)}$, $m_{k}^{(k)}$:
\begin{equation}\label{ZhS-eq41}
m_{k-1}^{(k)}=h_k\int\limits_0^1 a_{k-1}(t)\,dt, \quad m_{k}^{(k)}=h_k\int\limits_0^1 a_{k}(t)\,dt, \quad k=2,\dots,n.
\end{equation}

3. Taking into account the conditions (\ref{ZhS-eq15}) we determine:
\begin{equation}\label{ZhS-eq42}
y_{k-1}=x_{k-1}+a_0^{-1}(m_{k-1}-m_{k-1}^{(k-1)}-m_{k-1}^{(k)}), \quad x_k=y_{k-1}+h_k, \quad k=2,\dots,n,
\end{equation}
\begin{equation*}
m_{1}^{(1)}\equiv 0, \quad x_1=0, \quad y_n=L.
\end{equation*}

Note that failure inequalities
\begin{equation*}
m_{k-1}-m_{k-1}^{(k-1)}-m_{k-1}^{(k)}>0, \quad \quad k=2,\dots,n,
\end{equation*}
means that the parameter $\lambda$ is not chosen large enough.

\section{Examples of approximation}\label{ZhS-10}

We restrict the consideration by the case when
\begin{equation}\label{ZhS-eq43}
\mu_k=\mu, \quad \delta_k=\delta, \quad k=1,...,n.
\end{equation}
It is easy to get:
\begin{equation}\label{ZhS-eq44}
\int\limits_0^1 \Phi_k(t)\,dt=-
\frac{1}{\mu}
\int\limits_0^1 \left. \frac{\varphi'_{k-1}\varphi'_k}{\varphi'_k \varphi''_{k-1}-  \varphi'_{k-1} \varphi''_k} \right|_{\psi=\psi^{(k)}(t)}\,dt
\end{equation}
or
\begin{equation}\label{ZhS-eq45}
\int\limits_0^1 \Phi_k(t)\,dt=-
\frac{1}{\mu}\int\limits_{\psi_{k-1}}^{\psi_k}
\left.
\frac{\varphi'_{k-1}(\psi)\varphi'_k(\psi)}
{\varphi'_k(\psi) \varphi''_{k-1}(\psi) -  \varphi'_{k-1}(\psi) \varphi''_k(\psi)}
\cdot
\frac{1}{\displaystyle\frac{d\psi^{(k)}(t)}{dt}}
\right|_{t=t(\psi^{(k)})}\!\! d\psi,
\end{equation}
where the $t=t(\psi)$ is inverse function of the function $\psi=\psi^{(k)}(t)$. The inverse function exists because $\psi^{(k)}(t)$ is monotonic function.

\subsection{The linear function}\label{ZhS-10.1}

The simplest choice of $\psi^{(k)}(t)$ and, perhaps, not the best, is the linear function:
\begin{equation}\label{ZhS-eq46}
\psi^{(k)}(t)=(1-t)\psi_{k-1} + t\psi_k=\psi_{k-1}-t\Delta\psi_k,\quad
\Delta\psi_k=\psi_{k-1}-\psi_k>0.
\end{equation}

In this case the integral in (\ref{ZhS-eq45}) calculates easily. Using (\ref{ZhS-eq38}) we obtain:
\begin{equation}\label{ZhS-eq47}
h_k=
\frac{2a_0\mu\Delta\psi_k \sinh\Delta\psi_k}
{\lambda(\Delta\psi_k \cosh\Delta\psi_k- \sinh\Delta\psi_k)}.
\end{equation}

In the case of (\ref{ZhS-eq43}) for integrals in (\ref{ZhS-eq41}) we get:
\begin{equation}\label{ZhS-eq48}
m_{k-1}^{(k)}=m_{k}^{(k)}=\frac{1}{2} h_k a_0, \quad k=2,\dots,n.
\end{equation}
Note that the formula (\ref{ZhS-eq48}) will be valid always, if $\psi^{(k)}(t)$ be odd respect to $t=1/2$.

The disadvantage of the choice $\psi^{(k)}(t)$ as the linear function, in particular, is the presence of large magnitude discontinuities of the derivative at the points $x_k$, $y_k$. In the case of (\ref{ZhS-eq46}) gap derivatives, for example, at the point $x=x_k$ is:
\begin{equation}\label{ZhS-eq49}
\psi'(x_k+0)-\psi'(x_k-0)=h_k^{-1}\Delta\psi_k=O(\lambda), \quad \Delta\psi_k=O(1).
\end{equation}

\subsection{The nonlinear function}\label{ZhS-10.2}

Other choice of $\psi^{(k)}(t)$ is a nonlinear function, for example,
\begin{equation}\label{ZhS-eq50}
\psi^{(k)}(t)=\frac{\psi_k+\psi_{k-1}}{2}+\frac{\psi_k-\psi_{k-1}}{2}\frac{ \tanh\beta_k\left(t-\frac{1}{2} \right)}{\tanh\beta_k/2},
\end{equation}
\begin{equation*}
\psi^{(k)}(t)\bigr|_{t=0}=\psi_{k-1}, \quad \psi^{(k)}(t)\bigr|_{t=1}=\psi_k,
\end{equation*}
\begin{equation*}
\frac{d\psi^{(k)}}{dt}=-\frac{\Delta\psi_k}{2} \frac{\beta_k}{\cosh^2\beta_k(t-1/2)\tanh\beta_k/2},
\end{equation*}
where $\beta_k>0$ are some parameters.

In this case, the gap  derivatives (compare with (\ref{ZhS-eq49})):
\begin{equation}\label{ZhS-eq51}
\psi'(x_k+0)-\psi'(x_k-0)=
\frac{\Delta\psi_k\beta_k}{h_k\sinh\beta_k}<\frac{\Delta\psi_k}{h_k}, \quad \Delta\psi_k=\psi_{k-1}-\psi_k>0.
\end{equation}

The value of $h_k$ is determined by the formula (\ref{ZhS-eq38})
\begin{equation}\label{ZhS-eq52}
h_k=\frac{a_0}{\lambda\int_0^1 \Phi_k(t)\,dt},
\end{equation}
where
\begin{equation*}
 \int\limits_0^1 \Phi_k(t)\,dt=\frac{2\tanh\frac{\beta_k}{2}}{\mu\beta_k \Delta\psi_k \sinh\Delta\psi_k}
 \int\limits_{\psi_{k-1}}^{\psi_k}\frac{\sinh(\psi-\psi_k) \sinh(\psi-\psi_{k-1})d\psi}{1-\left[\frac{2\psi-\psi_{k-1}-\psi_k}{\Delta\psi_k} \right]^2 \tanh^2\frac{\beta_k}{2}}.
\end{equation*}
Value of $m_{k-1}^{(k)}$, $m_{k}^{(k)}$ are again determined by the formula (\ref{ZhS-eq48}).

Note that the result is weakly depends on the type of function
$\psi^{(k)}$. We mean that $h_k\to 0$ at $\lambda\to \infty$ for almost all monotonic decreasing function $\psi{(k)}$ is satisfying to (\ref{ZhS-eq21}).  	
Calculating the limit as $\lambda$ tending to infinity we get $h_k=0$ and
\begin{equation} \label{ZhS-eq53}
a_k(x)= \left\{
\begin{array}{ll}
0, & x\leqslant x_k, \\
a_0, & x_k \leqslant x\leqslant y_k, \\
0, & y_k \leqslant x,
\end{array}
\right.
\quad
\psi(x)=\psi_k,\quad x_k \leqslant x\leqslant y_k, \\
\quad k=1,\dots,n,
\end{equation}
Although, almost all approximation constructed in accordance with the algorithm give (\ref{ZhS-eq53}), the results for moderate values of $\lambda$ can be used to construct approximate weak solutions.

\section{Other way of the approximating functions selecting}\label{ZhS-11}

Here, we specify the approximation other than (\ref{ZhS-eq18}), (\ref{ZhS-eq19}).
For simplicity we restrict the consideration by the case when the parameters satisfy to (\ref{ZhS-eq18}), (\ref{ZhS-eq43}).

\begin{figure}[H]
\centering
\includegraphics[scale=0.85]{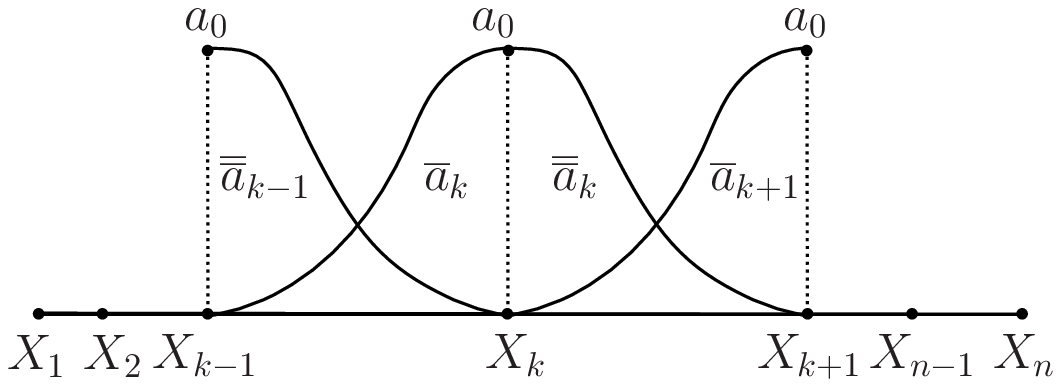}\\
\includegraphics[scale=0.85]{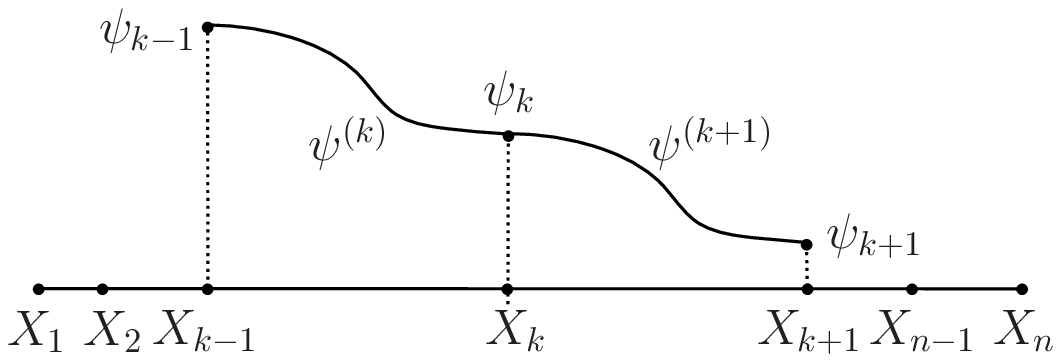}\\
\caption{Scheme of approximation}
\label{Ris2}
\end{figure}

We define the partition of interval $[0,L]$ by the set of points (see Fig.~\ref{Ris2})
\begin{equation*}
0=X_1 < X_2 < \dots < X_{k-1}< X_k < X_{k+1}<\dots < X_{n-1} < X_n=L.
\end{equation*}

To construct the solution of problem (\ref{ZhS-eq13})--(\ref{ZhS-eq16}) we use the approximation (see, Fig.~\ref{Ris2}).
\begin{equation} \label{ZhS-eq54}
 a_k(x)= \left\{
 \begin{array}{ll}
 0,                            &  x\leqslant X_{k-1}, \\
 \overline{a}_k(x),            &  X_{k-1} \leqslant  x\leqslant X_{k}, \\
 \overline{\overline{a}}_k(x), &  X_{k} \leqslant  x\leqslant X_{k+1}, \\
 0,                            &  X_{k+1} \leqslant x,
  \end{array}
        \right.
        \quad k=1,\dots,n,
\end{equation}
\begin{equation} \label{ZhS-eq55}
 \psi(x)= \psi^{(k)}(x),\quad  X_{k-1} \leqslant  x\leqslant X_{k},
          \quad k=2,\dots,n.
\end{equation}
Here, as before, $\overline{a}_k(x)$, $\overline{\overline{a}}_k(x)$, $\psi^{(k)}(x)$, $\psi^{(k+1)}(x)$ are functions which smooth at appropriate intervals and satisfying to continuity conditions:
\begin{equation}\label{ZhS-eq56}
 \overline{a}_k(X_{k-1})=0, \quad  \overline{a}_k(X_k)=a_0, \quad
 \overline{\overline{a}}_k(X_k)=a_0, \quad  \overline{\overline{a}}_k(X_{k+1})=0,
\end{equation}
\begin{equation}\label{ZhS-eq57}
 \psi^{(k)}(X_{k-1})=\psi_{k-1}, \quad  \psi^{(k)}(X_k)=\psi_k,
\end{equation}
\begin{equation*}
 \psi^{(k+1)}(X_{k})=\psi_k, \quad  \psi^{(k+1)}(X_{k+1})=\psi_{k+1}.
\end{equation*}

Further, we repeat almost verbatim the reasoning of the sections \ref{ZhS-06}--\ref{ZhS-08}. We consider the integrals on the interval $[X_{k-1},X_k]$  (symbols `overline' is omitted)
\begin{equation}
Q_k =\int\limits_{X_{k-1}}^{X_k}F_k(a;\psi^{(k)})V_k\,dx,
\label{ZhS-eq58}
\end{equation}
where $F_k(a;\psi^{(k)})$ is defined by (\ref{ZhS-eq23}) and has the form (see (\ref{ZhS-eq28}))
\begin{equation}\label{ZhS-eq59}
F_k(a;\psi^{(k)})=
-\frac{da_k}{dx} +
\frac{\lambda a_k \theta_k(\psi^{(k)})\sum\limits_{i=k-1}^k\!\!a_i\theta'_i(\psi^{(k)})}
{\sigma(a;\psi^{(k)})\sum\limits_{i=k-1}^k\!\!a_i
\left(
\theta^2_i(\psi^{(k)}) + \theta'_i(\psi^{(k)})
\right)}.
\end{equation}
As before, the concentration $a_k$ on the interval $[X_{k-1},X_k]$ is determined by the relations (\ref{ZhS-eq30}):
\begin{equation}\label{ZhS-eq60}
a_{k-1}=\frac{a_0 \theta_k(\psi^{k})}{ \theta_k(\psi^{k})- \theta_{k-1}(\psi^{k})},\quad
a_{k}(t)=-\frac{a_0 \theta_{k-1}(\psi^{k})}{\theta_k(\psi^{k})- \theta_{k-1}(\psi^{k})}.
\end{equation}

We assume that $\psi^{(k)}(x)$ is defined by the differential equation
\begin{equation}\label{ZhS-eq61}
\frac{d\psi^{(k)}}{dx}=-\frac{\lambda\sum\limits_{i=k-1}^k a_i\theta_i^2(\psi^{(k)})}
{\sigma(a;\psi^{(k)})\sum\limits_{i=k-1}^k a_i \left(\theta^2_i(\psi^{(k)}) + \theta'_i(\psi^{(k)})\right)}-\omega_k^2<0,
\end{equation}
where $\omega_k^2>0$ is some parameter.

At small $\omega^2$ the equation (\ref{ZhS-eq61})  is some perturbation of the equation (\ref{ZhS-eq12}) for the original problem
(\ref{ZhS-eq7})--(\ref{ZhS-eq11}). Choice of $\omega_k^2=0$, unfortunately, is impossible.
It is easy to check that if the approximation of (\ref{ZhS-eq54})--(\ref{ZhS-eq57}), (\ref{ZhS-eq60}) is
chosen then integrals in (\ref{ZhS-eq15}) are singularity, as
\begin{equation}\label{ZhS-eq62}
\sum\limits_{i=k-1}^k a_i\theta_i^2(\psi^{(k)})=0, \quad
\psi^{(k)}=\psi_{k-1},\psi_{k}.
\end{equation}

We add the condition (\ref{ZhS-eq57}) to the equation (\ref{ZhS-eq61}).
One of these conditions is required for Cauchy problem and other condition is determined the difference $(X_k-X_{k-1})$.


Taking into account that functions $a_{k-1}$ and $a_{k}$ depend only on the $\psi_{k}$ we rewrite (\ref{ZhS-eq59}) in the form
\begin{equation}\label{ZhS-eq63}
F_k(a;\psi^{(k)})=
-\frac{da_k}{d\psi^{(k)}}\frac{d\psi^{(k)}}{dx} +
\frac{\lambda a_k \theta_k(\psi^{(k)})\sum\limits_{i=k-1}^k\!\!a_i\theta'_i(\psi^{(k)})}
{\sigma(a;\psi^{(k)})\sum\limits_{i=k-1}^k\!\!a_i
\left(
\theta^2_i(\psi^{(k)}) + \theta'_i(\psi^{(k)})
\right)}.
\end{equation}
We obtain the derivative $da_k/d\psi^{(k)}$ taking into account (\ref{ZhS-eq60}) and differentiating the system (\ref{ZhS-eq29}) with respect to $\psi$.  Substituting $d\psi^{(k)}/dx$ from (\ref{ZhS-eq61}) to (\ref{ZhS-eq63}) after simple transformations we have
\begin{equation}\label{ZhS-eq64}
F_k(a;\psi^{(k)})=
-\omega^2
\frac{\sum\limits_{i=k-1}^k a_i \theta'_i(\psi^{(k)})
}{\theta_k(\psi^{(k)})-\theta_{k-1}(\psi^{(k)})}.
\end{equation}

\subsection{The choice of the parameter $\omega_k^2$}\label{ZhS-11.1}

It is possible to choose the parameter $\omega_k^2\to 0$ at $\lambda\to\infty$. In this case the estimation $F_k(a;\psi^{(k)})=O(\omega_k^2)$ means that the approximation (\ref{ZhS-eq54})--(\ref{ZhS-eq57}), (\ref{ZhS-eq60}) and the solution $\psi^{(k)}$ of the differential equation (\ref{ZhS-eq61}) are the weak solution of (\ref{ZhS-eq13})--(\ref{ZhS-eq16}).

We restrict the consideration by the case when
\begin{equation}\label{ZhS-eq65}
\psi_{k-1}-\psi_{k}=\Delta\psi, \quad k=2,\dots,n, \quad M_1=M_n=\frac{M}{2}, \quad M_k=M, \quad k=2,\dots,n-1,
\end{equation}
\begin{equation*}
\mu_k=\mu, \quad \delta_k=\delta, \quad k=1,...,n.
\end{equation*}

We introduce the notation for the right side of the differential equation (\ref{ZhS-eq61})
\begin{equation}\label{ZhS-eq66}
R_k(\psi^{(k)})=\frac{\sum\limits_{i=k-1}^k a_i\theta_i^2(\psi^{(k)})}
{\sigma(a;\psi^{(k)})\sum\limits_{i=k-1}^k a_i \left(\theta^2_i(\psi^{(k)}) + \theta'_i(\psi^{(k)})\right)}.
\end{equation}
Then, the length of the interval $[X_{k-1}-X_k]$ has the form
\begin{equation}\label{ZhS-eq67}
H_k=X_{k}-X_{k-1}=\int\limits_{X_{k-1}}^{X_{k}}dx=
\int\limits_{\psi_{k-1}}^{\psi_{k}}\frac{d\psi}{\psi'(x)}=
-\int\limits_{\psi_{k-1}}^{\psi_{k}}\frac{d\psi}{\lambda R_k(\psi)+\omega_k^2}.
\end{equation}

It is easy to show that in the case (\ref{ZhS-eq65}) all the parameters $\omega_k^2=\omega^2$ and the distribution of the concentrations of $a_k(x)$, $\psi^{(k)}(x)$, and $R_k(\psi^{(k)})$ are symmetric functions with respect to bisecting point of a segment $[X_{k-1}-X_k]$.

The values $X_k$ are defined by relations:
\begin{equation}\label{ZhS-eq68}
X_{2}-X_{1}=\frac{1}{2}H, \quad X_{k}-X_{k-1}=H, \quad k=3,\dots,n-1, \quad X_{n}-X_{n-1}=\frac{1}{2}H,
\end{equation}
where
\begin{equation*}
H=\int\limits_{\psi_{k}}^{\psi_{k-1}}\frac{d\psi}
{\lambda R_k(\psi)+\omega^2}.
\end{equation*}

Using the symmetric properties of function $R_k(\psi)$ one can get the asymptotic relation at $\omega^2/\lambda \to 0$ (see detail in Appendix~\ref{App:ZhS-1}):
\begin{equation}\label{ZhS-eq69}
\frac12 \lambda H=-\frac{1}{R'_k(\psi_{k})}\ln\frac{\omega_0^2}{R_k(\psi_*)} + \frac{\ln W}{R'_k(\psi_{k})} +O(\omega_0^2 \ln\omega_0^2),
\end{equation}
\begin{equation*}
\psi_*=\frac12 (\psi_{k-1}+\psi_{k}), \quad \omega_0^2=\frac{\omega^2}{\lambda},
\end{equation*}

where $W$ is constant that does not depend on $\lambda$ and $\omega_0^2$.

Then, we have
\begin{equation}\label{ZhS-eq70}
\omega^2=\lambda W R_k(\psi_*) \exp\left(-\frac{1}{2}\lambda H R'_k(\psi_k)\right)\to 0, \quad
|Q_k|=O(\omega^2), \quad \lambda\to \infty.
\end{equation}
This estimate means that the approximation
(\ref{ZhS-eq54})--(\ref{ZhS-eq57}), (\ref{ZhS-eq60}), (\ref{ZhS-eq61}) is a weak solution of the problem.

Note, the derivative $d\psi/dx$ is continuous at the points $X_k$ when the parameters satisfy (\ref{ZhS-eq65}). In the general case the gap derivatives, obviously, would be equal ($\omega^2_k-\omega^2_{k-1}$).

For practical accurate calculations we should solve equation (\ref{ZhS-eq67}) relative to $\omega^2$ at given value $H_k$, which
for the case (\ref{ZhS-eq65}) is defined by the conditions (see (\ref{ZhS-eq15}))
\begin{equation}\label{ZhS-eq71}
\int\limits_{X_{k-1}}^{X_k}a_k\,dx=\frac{1}{2}a_0 H_k=\frac{1}{2}M_k.
\end{equation}

\section{Weak solutions approximation at moderate parameter}\label{ZhS-12}

Despite the fact that the main result for the weak solution of the problem is obtained for $\lambda \to \infty$ it can be efficiently used at moderate values of the parameter $\lambda$. In Appendix~\ref{App:ZhS-1} the comparison of the numerical solution of the equation (\ref{ZhS-eq67}) and asymptotic formula (\ref{ZhS-eq70}) is presented.

To demonstrate the method of the weak solution construction we choose the following parameters:
\begin{equation}\label{ZhS-eq72}
\psi_1=5,\quad \Delta\psi=1, \quad n=11, \quad \mu=1, \quad \delta=3, \quad
M=0.1, \quad L=1, \quad \lambda=200,
\end{equation}
\begin{equation*}
\mu_1=\dots=\mu_{11}=\mu, \quad \delta_1=\dots=\delta_{11}=\delta,
\end{equation*}
\begin{equation*}
M_1=\frac12 M, \quad M_2=\dots=M_{10}=M, \quad M_{11}=\frac12 M.
\end{equation*}
Using (\ref{ZhS-eq65}), (\ref{ZhS-eq68}), (\ref{ZhS-eq11}) we have
\begin{equation}\label{ZhS-eq73}
H_1=\dots=H_{10}=H=0.1, \quad  a_0=1.
\end{equation}
At $\lambda=200$ we get (see Appendix~\ref{App:ZhS-1}, Tab.~\ref{tab:table1})
\begin{equation*}
W=5.968, \quad R'_k(\psi_k)=1.035, \quad R_k(\psi_*)=0.227.
\end{equation*}
Using formula (\ref{ZhS-eq70}) (or (\ref{ZhSeq-A9})) we have
\begin{equation*}
\omega_k^2=\omega^2=0.00868.
\end{equation*}
We solve the Cauchy problem (\ref{ZhS-eq61}) on the interval $[X_{k-1},X_k]$
\begin{equation}\label{ZhS-eq74}
\frac{d\psi^{(k)}}{dx}=-\frac{\lambda\sum\limits_{i=k-1}^k a_i\theta_i^2(\psi^{(k)})}
{\sigma(a;\psi^{(k)})\sum\limits_{i=k-1}^k a_i \left(\theta^2_i(\psi^{(k)}) + \theta'_i(\psi^{(k)})\right)}-\omega_k^2<0,
\end{equation}
\begin{equation*}
\psi^{(k)}(X_{k-1})=\psi_{k-1}
\end{equation*}
and simultaneously determine the concentration $a_k$ with the help of formulae (\ref{ZhS-eq61})
\begin{equation}\label{ZhS-eq75}
a_{k-1}=\frac{a_0 \theta_k(\psi^{k})}{ \theta_k(\psi^{k})- \theta_{k-1}(\psi^{k})},\quad
a_{k}(t)=-\frac{a_0 \theta_{k-1}(\psi^{k})}{\theta_k(\psi^{k})- \theta_{k-1}(\psi^{k})}.
\end{equation}
Note that in the case (\ref{ZhS-eq72}), (\ref{ZhS-eq73}) it is enough to solve the initial value problem on any one interval and then to continue solution on subsequent intervals `periodically'.

\begin{figure}[H]
\centering
\includegraphics[scale=0.9]{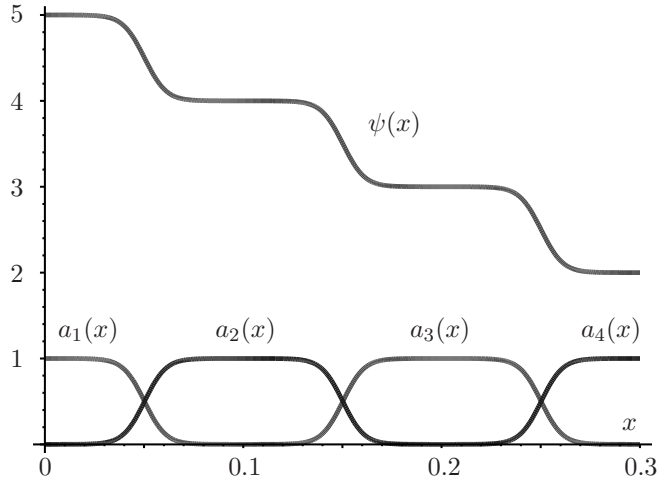}\\
\caption{The distribution of the concentrations $a_k(x)$ and acidity function $\psi(x)$. The fragment of approximation at $\lambda=200$}
\label{Ris3}
\end{figure}

On Fig.~\ref{Ris3} the results of numerical integration are shown.

\subsection{Comparison between the weak solution and the solution of the original problem}\label{ZhS-12.1}

Algorithm of the numerical integration of the original problem (\ref{ZhS-eq7})--(\ref {ZhS-eq10}) is described in \cite{Averkov} and its modification presented in
\cite{Part2,SakhVladZhuk}. We compare the numerical solution with the weak solution for the following parameters (see also Appendix~\ref{App:ZhS-1}):
\begin{equation*}
\psi_1=5, \quad \psi_{k-1}-\psi_{k}=\Delta\psi=1, \quad \mu_k=\mu=1, \quad \delta_k=\delta=15, \quad a_0=1, \quad H_k=0.25.
\end{equation*}

\begin{figure}[H]
\centering
\includegraphics[scale=0.9]{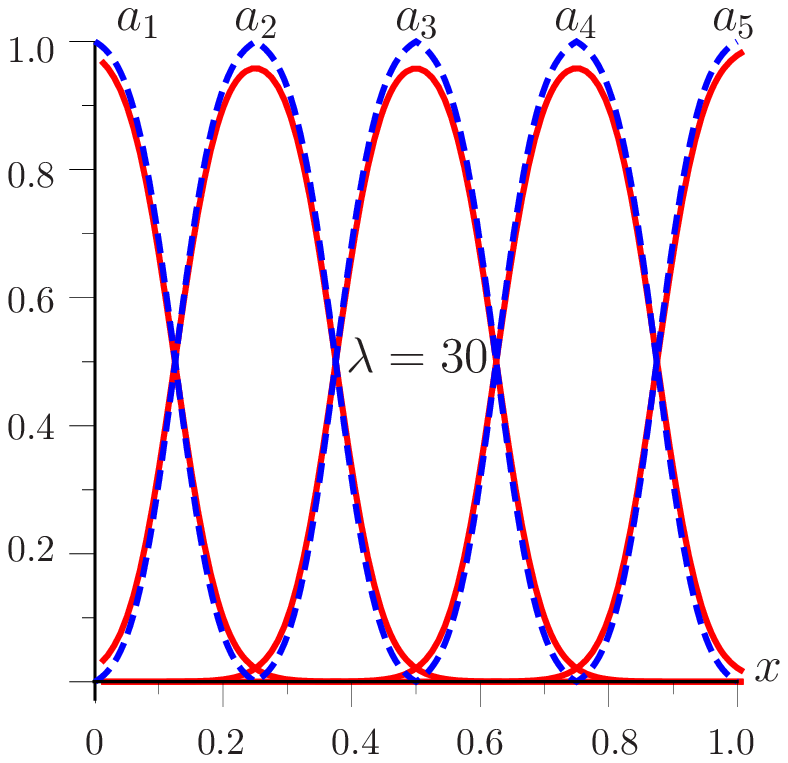}
\includegraphics[scale=0.9]{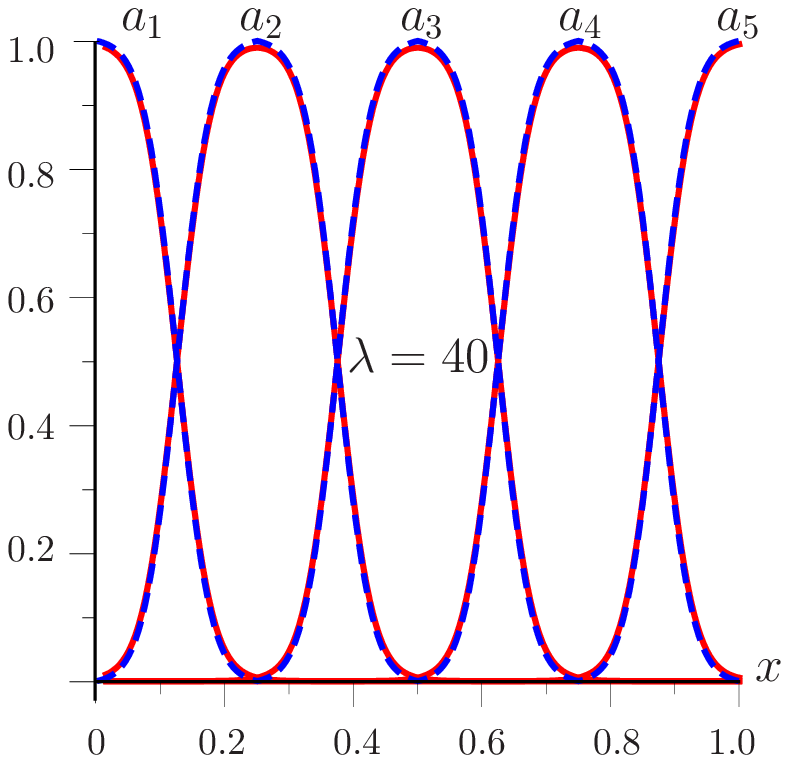}\\
\caption{The distribution of the concentration at $\delta=15$. Comparison between the weak solution and the numerical solution.
$\lambda=30$, $U_0=15.063$ ($I_*=7.439   \,\,\mu\textrm{A}$,  $U_*=15.213        \,\,\textrm{V}$);
$\lambda=40$, $U_0=15.290$ ($I_*=9.919   \,\,\mu\textrm{A}$,  $U_*=15.442        \,\,\textrm{V}$)
}
\label{Ris4}
\end{figure}

\begin{figure}[H]
\centering
\includegraphics[scale=0.9]{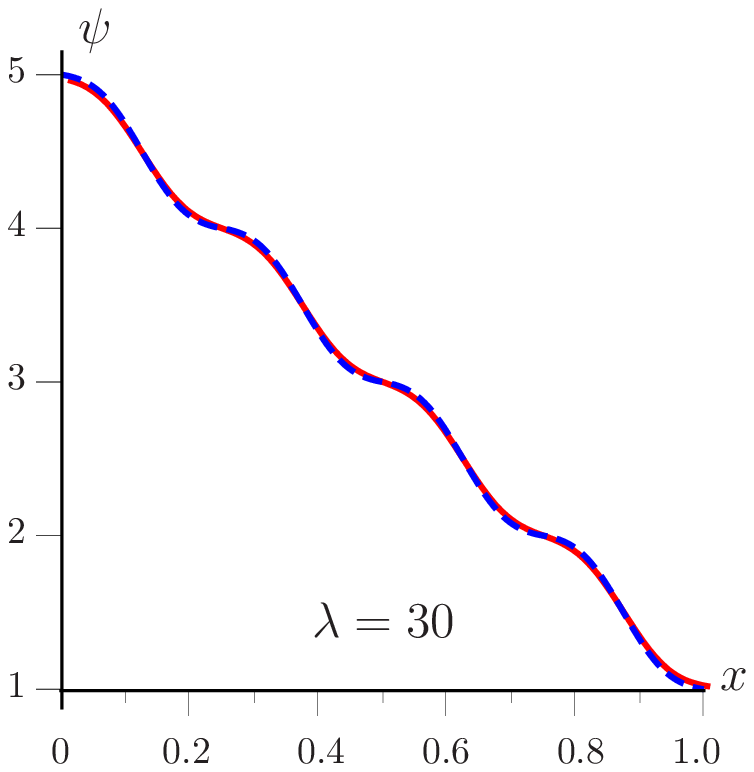}
\includegraphics[scale=0.9]{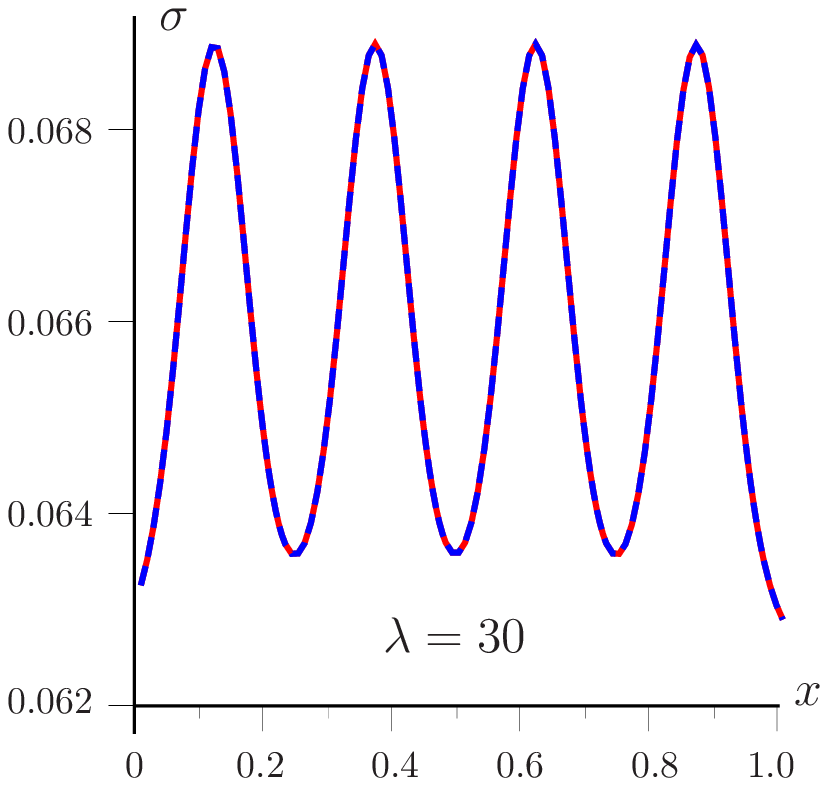}\\
\caption{The distribution of the acidity $\psi(x)$ and conductivity $\sigma(x)$ at $\delta=15$. Comparison between the weak solution and the numerical solution}
\label{Ris5}
\end{figure}

On Figs.~\ref{Ris4}, \ref{Ris5} the results of calculation are shown.

Starting from parameters $\lambda=30$ we have a good agreement between the weak solution (\ref{ZhS-eq74}), (\ref{ZhS-eq75}) and the solution of original problem
(\ref{ZhS-eq7})--(\ref {ZhS-eq10}).

\section{Conclusion}\label{ZhS-13}

Described technique of constructing the weak solutions for the original problem is quite specific. Success is primarily due to the fact that the presence of a small parameter at higher derivatives and turning points dictates specific structure of solution. For large values of the parameter $\lambda$ the functions $a_k$ are almost completely focused at certain intervals. Moreover, these functions quickly and exponentially decrease outside of own intervals (see (\ref{ZhS-eq17}) and Sec.\,\ref{ZhS-05}, \ref{ZhS-11}). It allows to split a system of $n$ equations on a separate subsystems containing only two equations. However, this involve the additional difficulties. The problem of determination of the acidity function becomes singular. The removing of this singularity is possible by the introduction of some perturbation of the problem (see (\ref{ZhS-eq61}) and Sec.\,\ref{ZhS-11}).

One of the most interesting result is the fact that at $\lambda=\infty$ a generalized solution of the original problem is occurred (see (\ref{ZhS-eq53})). At moderate values of the parameter $\lambda$ approximation of a weak solution is actually the asymptotic of the original problem solution. Confirmation of this fact is a good coincidence of the weak solution of the problem and the numerical solution of the problem.
In more detail the process of separation will be described in \cite{Part3} which gives the solution of non-stationary problem.

\begin{acknowledgments}
This research is partially supported by Russian Foundation for Basic Research (grants 10-05-00646 and 10-01-00452),
Ministry of Education and Science of the Russian Federation
(programme `Development of the research potential of the high school', contracts  14.A18.21.0873,  8832 and grant 1.5139.2011).
The authors are grateful to N.\,M.~Zhukova for reviewing the translated text into English.

\end{acknowledgments}


\appendix

\section{Asymptotic evaluation of integral (\ref{ZhS-eq67})}\label{App:ZhS-1}

Using the symmetric properties of function $R_k(\psi)$ we can get:
\begin{equation}\label{ZhSeq-A1}
\frac12 \lambda H_k=
\int\limits_{\psi_{*}}^{\psi_{k-1}}\frac{d\psi}{R_k(\psi)+\omega_0^2}, \quad \omega_0^2=\frac{\omega^2}{\lambda}, \quad \psi_{*}=\frac12(\psi_{k}+\psi_{k-1}).
\end{equation}
We recall that
\begin{equation}\label{ZhSeq-A2}
R_k(\psi_{k-1})=0, \quad R'_k(\psi_{k-1})<0, \quad R'_k(\psi_{*})=0,  \quad R_k(\psi_{*})>0.
\end{equation}
We change variables:
\begin{equation}\label{ZhSeq-A3}
R_k(\psi)=\tau,  \quad R_k(\psi_{*})=\tau_0 > 0, \quad \mathcal{F}(\tau)=\frac{1}{R'_k(\psi)}.
\end{equation}
Then, we can rewrite (\ref{ZhSeq-A1}) in the form:
\begin{equation}\label{ZhSeq-A4}
\frac12 \lambda H_k=
\int\limits_{R_k(\psi_{*})=\tau_0}^{R_k(\psi_{k-1})=0}\frac{d\tau}{R'_k(\psi)(\tau+\omega_0^2)}=
\int\limits_{\tau_0}^{0}\frac{\mathcal{F}(\tau) d\tau}{\tau+\omega_0^2}.
\end{equation}

We note that integrand has the integrable singularity in vicinity of point $\tau=\tau_0$. Actually, using  (\ref{ZhSeq-A3}) we have:
\begin{equation}\label{ZhSeq-A5}
R_k(\psi)=\tau=R_k(\psi_*)+\frac12R''_k(\psi_*)(\tau-\tau_0)^2+\cdots=\tau_0+\frac12R''_k(\psi_*)(\psi-\psi_*)^2+\cdots,
\end{equation}
\begin{equation*}
(\psi-\psi_*)\approx \left(\frac{2(\tau-\tau_0)}{R''_k(\psi_*)}\right)^{1/2}, \quad \tau<\tau_0, \quad R''_k(\psi_*)<0,
\end{equation*}
\begin{equation*}
R'_k(\psi)=R''_k(\psi_*)(\psi-\psi_*)+\cdots\approx  R''_k(\psi_*) \left(\frac{2(\tau-\tau_0)}{R''_k(\psi_*)}\right)^{1/2},
\end{equation*}
\begin{equation*}
\mathcal{F}(\tau) =O((\tau_0-\tau)^{-1/2}), \quad \tau \to \tau_0.
\end{equation*}
Further, we present (\ref{ZhSeq-A4}) in the form (we construct the asymptotic following \cite{Fedoruk}):
\begin{equation}\label{ZhSeq-A6}
\frac12 \lambda H_k=
+\mathcal{F}(-\omega_0^2)\ln\frac{\omega_0^2}{\tau_0+\omega_0^2}
\int\limits_{\tau_0}^{0}\frac{\mathcal{F}(\tau) - \mathcal{F}(-\omega_0^2)}{\tau+\omega_0^2}\,d\tau
\end{equation}
We keep principal terms only and write:
\begin{equation}\label{ZhSeq-A7}
\frac12 \lambda H_k=
\mathcal{F}(0)\ln\frac{\omega_0^2}{\tau_0}+
\int\limits_{\tau_0}^{0}\frac{\mathcal{F}(\tau) - \mathcal{F}(0)}{\tau}\,d\tau+O(\omega_0^2 \ln \omega_0^2).
\end{equation}
Taking into account that $R'_k(\psi_{k})=-R'_k(\psi_{k-1})>0$ we get:
\begin{equation}\label{ZhSeq-A8}
\frac12 \lambda H_k=
-\frac{1}{R'_k(\psi_{k})}\ln\frac{\omega_0^2}{R_k(\psi_{*})}+
\frac{\ln W}{R'_k(\psi_{k})}+O(\omega_0^2 \ln \omega_0^2),
\end{equation}
where
\begin{equation}\label{ZhSeq-A9}
W=\exp
\left\{
R'_k(\psi_{k})\int\limits_{\tau_0}^{0}\frac{\mathcal{F}(\tau) - \mathcal{F}(0)}{\tau}\,d\tau
\right\}.
\end{equation}
Finally, we have
\begin{equation}\label{ZhSeq-A10}
\omega^2=\lambda W R_k(\psi_*) \exp\left(-\frac{1}{2}\lambda H_k R'_k(\psi_k)\right)\to 0,  \quad \lambda\to \infty.
\end{equation}
Integral in formula (\ref{ZhSeq-A9}) has not singularity and can be calculated by numerical methods.

Other way for calculating integral is the application of the Taylor series:
\begin{equation}\label{ZhSeq-A11}
\int\limits_{\tau_0}^{0}\frac{\mathcal{F}(\tau) - \mathcal{F}(0)}{\tau}\,d\tau=
-\sum\limits_{m=1}^{\infty}\frac{\mathcal{F}^{(m)}(0)}{m\, m!}\tau_0^m,
\end{equation}
\begin{equation*}
\mathcal{F}^{(m)}(0)=\left.\frac{d^m\mathcal{F}(\tau)}{d\tau^m}\right|_{\tau=0}=
\left.
\left(
\frac{1}{R'_k(\psi)}
\frac{d}{d \psi}
\right)^m \frac{1}{R'_k(\psi)}
\right|_{\psi=\psi_{k-1}}.
\end{equation*}

\begin{figure}[H]
\centering
\includegraphics[scale=0.72]{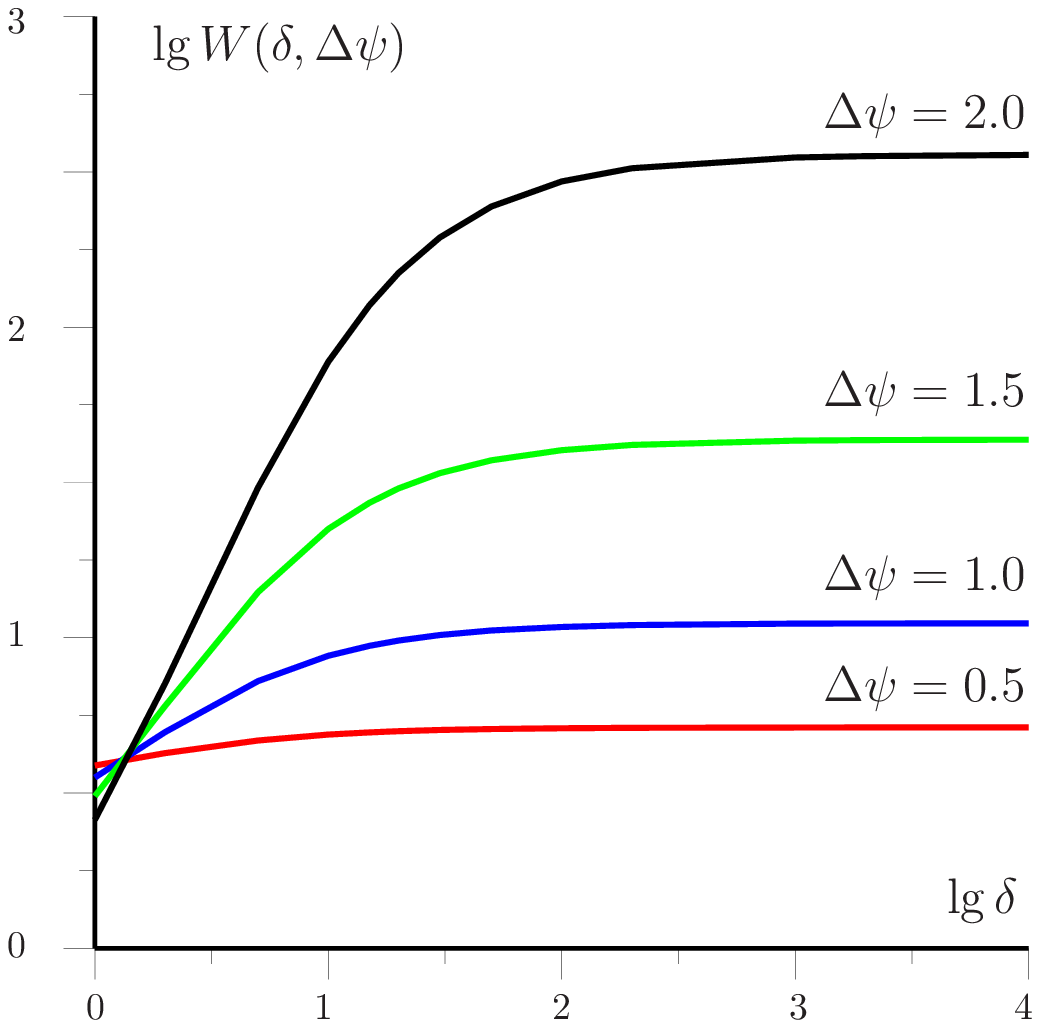}
\includegraphics[scale=0.72]{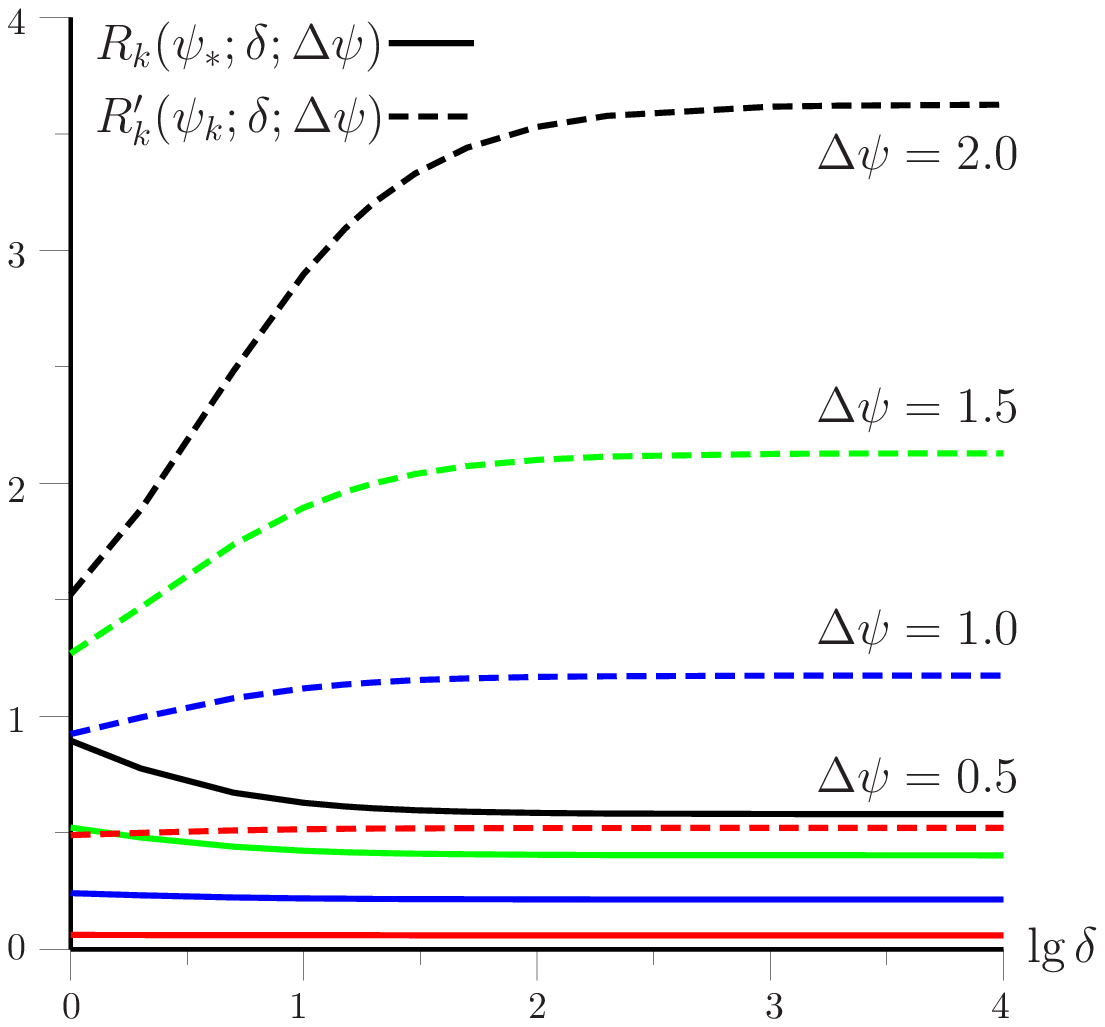}\\
\caption{Dependences $W(\delta,\Delta\psi)$, $R_k(\psi_*;\delta;\Delta\psi)$,  $R'_k(\psi_k;\delta;\Delta\psi)$ on $\delta$. See table~\ref{tab:table1}}
\label{Ris6}
\end{figure}

The results of calculation presented in Tab.~\ref{tab:table1} and on Fig.~\ref{Ris6} for the following parameters:
\begin{equation*}
\psi_{k-1}-\psi_{k}=\Delta\psi, \quad \mu_k=\mu, \quad \delta_k=\delta.
\end{equation*}
Note, that $W$, $\mu a_0 R'_k(\psi_k)$, $ \mu a_0 R_k(\psi_*)$ almost do not depend on $\delta$ starting from $\delta\approx 100$.
In particular, this means that one can assume $\delta_k=\delta > 100$.

\begin{table}[H]
\caption{\label{tab:table1} $W(\delta,\Delta\psi)$, $R'_k(\psi_k)$, $R_k(\psi_*)$}
\begin{ruledtabular}
\begin{tabular}{c|cccc|cccc|cccc}
$\delta$          & $W$       &          &          &           & $\mu a_0 R'_k(\psi_k)$ &     &         &        & $ \mu a_0 R_k(\psi_*)$ &     &         & \\
\hline
      &$\Delta\psi=0.5$       & $1.0$    & $1.5$    & $2.0$     &  $ 0.5$    & $1.0$   & $1.5$   & $2.0$  &  $0.5$    & $1.0$   & $1.5$   & $2.0$\\
\hline
$1$               &  $3.878$  & $3.547$  & $3.090$  & $2.592$   &  $ 0.491$  & $0.924$ & $1.270$ & $1.523$&  $0.062$  & $0.241$ & $0.522$ & $0.895$\\
$2$               &  $4.249$  & $4.969$  & $6.029$  & $7.121$   &  $ 0.500$  & $0.995$ & $1.468$ & $1.888$&  $0.061$  & $0.231$ & $0.479$ & $0.776$\\
$3$               &  $4.452$  & $5.968$  & $8.947$  & $13.641$  &  $ 0.505$  & $1.035$ & $1.591$ & $2.145$&  $0.061$  & $0.227$ & $0.459$ & $0.722$\\
$5$               &  $4.668$  & $7.249$  & $14.021$ & $30.464$  &  $ 0.510$  & $1.078$ & $1.738$ & $2.483$&  $0.061$  & $0.222$ & $0.440$ & $0.672$\\
$10$              &  $4.876$  & $8.747$  & $22.404$ & $77.309$  &  $ 0.515$  & $1.120$ & $1.896$ & $2.899$&  $0.060$  & $0.218$ & $0.423$ & $0.629$\\
$15$              &  $4.958$  & $9.414$  & $27.196$ & $117.581$ &  $ 0.517$  & $1.137$ & $1.963$ & $3.093$&  $0.060$  & $0.217$ & $0.417$ & $0.613$\\
$20$              &  $5.001$  & $9.789$  & $30.239$ & $149.349$ &  $ 0.518$  & $1.146$ & $2.000$ & $3.205$&  $0.060$  & $0.216$ & $0.414$ & $0.605$\\
$30$              &  $5.046$  & $10.198$ & $33.855$ & $194.272$ &  $ 0.519$  & $1.155$ & $2.040$ & $3.330$&  $0.060$  & $0.215$ & $0.410$ & $0.597$\\
$50$              &  $5.084$  & $10.552$ & $37.260$ & $244.555$ &  $ 0.520$  & $1.163$ & $2.074$ & $3.441$&  $0.060$  & $0.215$ & $0.408$ & $0.590$\\
$100$             &  $5.113$  & $10.834$ & $40.175$ & $294.483$ &  $ 0.520$  & $1.169$ & $2.101$ & $3.530$&  $0.060$  & $0.214$ & $0.406$ & $0.585$\\
$200$             &  $5.128$  & $10.981$ & $41.767$ & $324.653$ &  $ 0.521$  & $1.172$ & $2.115$ & $3.578$&  $0.060$  & $0.214$ & $0.404$ & $0.583$\\
$10^3$            &  $5.140$  & $11.102$ & $43.111$ & $351.837$ &  $ 0.521$  & $1.175$ & $2.126$ & $3.617$&  $0.060$  & $0.214$ & $0.404$ & $0.581$\\
$10^4$            &  $5.143$  & $11.129$ & $43.422$ & $358.369$ &  $ 0.521$  & $1.175$ & $2.129$ & $3.626$&  $0.060$  & $0.214$ & $0.403$ & $0.580$\\
\end{tabular}
\end{ruledtabular}
\end{table}

In Tab.~\ref{tab:table2} the numerical solution $\omega^2$ of the equation (\ref{ZhS-eq67}) and asymptotic values $\omega_a^2$ calculated by formula (\ref{ZhSeq-A10}) are presented for the following parameters:
\begin{equation*}
\psi_{k-1}-\psi_{k}=\Delta\psi=1, \quad \mu_k=\mu=1, \quad \delta_k=\delta=15, \quad a_0=1, \quad H_k=0.25.
\end{equation*}

\begin{table}[H]
\caption{\label{tab:table2} Dependence $\omega^2$ and its asymptotic value
$\omega_a^2$ on parameter $\lambda$ at $\delta=15$. See table~\ref{tab:table1}}
\begin{ruledtabular}
\begin{tabular}{ccccccccc}
$\lambda$ & $10$ & $20$ & $30$ & $40$ & $50$  & $60$ & $70$ & $80$\\
\hline
$\omega^2$                  & $2.579329$ & $1.439443$ & $0.663345$ & $0.249932$ & $0.080934$  & $0.024045$ & $0.006828$ & $ 0.001889$\\
$\omega_{a}^2$              & $4.929054$ & $2.380993$ & $0.862609$ & $0.277790$ & $0.083867$  & $0.024307$ & $0.006849$ & $ 0.001891$\\
$\omega_{a}^2/\omega^2$     & $1.910983$ & $1.654107$ & $1.300393$ & $1.111465$ & $1.036242$  & $1.010913$ & $1.003144$ & $ 1.000880$\\
\end{tabular}
\end{ruledtabular}
\end{table}

\section{Generalized solution of the stationary problem}\label{App:ZhS-5}

At $\lambda=\infty$ (or $\varepsilon=0$) it is easily to construct the generalized solution of the stationary problem
(\ref{ZhS-eq7})--(\ref{ZhS-eq11}). Assuming $\lambda=\infty$ we rewrite the problem in the following form
\begin{equation}\label{ZhSeq-E1}
a_k(x)\theta_k(\psi(x))=0, \quad
\left( \int\limits_0^L a_k(x)\theta_k(\psi(x))v(x)\,dx=0 \right), \quad  k=1,...,n,
\end{equation}
\begin{equation}\label{ZhSeq-E2}
   \int\limits_0^L a_k(x)\,dx = M_k,
\end{equation}
where $a_k(x)$, $\psi(x)$ are the generalized functions (distributions), $v(x)$ is a compact function.

Obviously, we have the following solution (compare with (\ref{ZhS-eq53}))
\begin{equation} \label{ZhSeq-E3}
a_k(x)= \left\{
\begin{array}{ll}
0, & x \leqslant \overline{x}_{k-1}, \\
a_0, & \overline{x}_{k-1} \leqslant x\leqslant \overline{x}_k, \\
0, & \overline{x}_k \leqslant x,
\end{array}
\right.
\quad
\psi(x)=\psi_k,\quad \overline{x}_{k-1} \leqslant x\leqslant \overline{x}_k, \\
\quad k=1,\dots,n,
\end{equation}
where
\begin{equation}\label{ZhSeq-E4}
   \overline{x}_0=0, \quad \overline{x}_k=\overline{x}_{k-1}+M_ka_0^{-1},  \quad k=1,\dots,n, \quad \overline{x}_n=L, \quad
   a_0=L^{-1}\sum\limits_{i=1}^n M_i
\end{equation}

\setlength{\bibsep}{4.0pt}


\begin{thebibliography}{99}

\small

\bibitem{ZhukovBabskiyYudovich}
\textsl{Babsky~V.\,G., Zhukov~M.\,Yu., Yudovich~V.\,I.} Mathematical theory of electrophoresis. Kiev: Naukova Dumka, 1983.

\bibitem{BabZhukYudE}
\textsl{Babsky~V.\,G., Zhukov~M.\,Yu., Yudovich~V.\,I.}
\emph{Mathematical theory of electrophoresis} (Plenum Publishing
Corporation, New York, 1989).


\bibitem{MosherSavilleThorman}
\textsl{Mosher~R.\,A., Saville~D.\,A., Thorman~W.} The Dynamics of
Electrophoresis. VCH Publishers, New York, 1992. 236\;p.


\bibitem{Righetti83}
\textsl{Righetti~P.\,G.} Isoelectric focusing: Theory, Methodology and Application. Elsevier Biomedical Press, New York--Oxford: Elsevier, 1983. 386~p.


\bibitem{Righetti90}
\textsl{Righetti~P.\,G.} Immobilized pH gradient: theory and methodology.
Laboratory techniques in biochemistry and molecular biology.
Elsevier Biomedical Press, Amsterdam, New York--Oxford: Elsevier,
1990. 397~p.

\bibitem{ZhRStoy2001}
\textsl{Stoyanov~A., Zhukov~M.\,Yu., Righetti ~P.\,G.}
The Proteome Revisited: Theory and practice of all relevant
electrophoretic steps // J. Chromatography. 2001. Vol.\,63
Elsevier, 2001. Chem. 572.6 R571 P967 2001. P.\,1--462.

\bibitem{Vesterberg1966}
Vesterberg~O., Svensson H. Isoelectric fractionation, analysis and characterization of ampholytes in natural pH gradients. IV. Further studies on the resolving power in
connection with the separation of myoglobins. Acta Chem. Scand., 1966, 20, P. 820--834.



\bibitem{Vesterberg1976}
Vesterberg~O. The carrier ampholytes. Isoelectric focusing. Acad. pres, New York-London. 1976. P. 53--76.


\bibitem{Haglund1971}
Haglund~H. Isoelectric focusing in pH gradients -- a technique for fractionation and characterization of ampholytes. Meth. Biochem. anal. 1971. ¹ 19. P. 1--104.
\bibitem{Svensson1961}
Svensson~H. Isoelectric fractionation, analysis, and characterization of ampholytes in
natural pH gradients. I. The differential equation of solute concentrations at a steady state
and its solution for simple cases. Acta chem. scand. 1961, 15, ¹ 2. P. 325--341.


\bibitem{Thormann2004}
Thormann~W., Huang~T., Pawliszyn~J., Mosher~R.~A. High-resolution computer simulation of the dynamics of isoelectric focusing of proteins. Electrophoresis. 2004, ¹~25. P. 324-337.

\bibitem{Thormann2006}
Thormann~W., Mosher~R.~A. High-resolution computer simulation of the dynamics of isoelectric focusing using carrier ampholytes: Focusing with concurrent electrophoretic mobilization is an isotachophoretic process. Research Article. Electrophoresis. 2006, No.~27. P. 968--983.

\bibitem{Averkov}
Averkov~A.\,N., Zhukov~M.\,Yu., Sakharova~L.\,V.  Calculation of the stationary $\rm{pH}$-gradient in aminoacid solution at large current density. Proc. IX International Conf.  `Modern problem of the continuum media', Rostov-on-Don, 2005. V.1. TsVVR Press, Rostov-on-Don.  P. 8--13.

\bibitem{Part2}
Sakharova~L.\,V., Shiryaeva ~E.\,V., Zhukov~M.\,Yu.
Mathematical Model of a pH-gradient Creation at~Isoelectrofocusing. Part II. Numerical Solution of the Stationary Problem.
arXiv:

\bibitem{Part3}
Shiryaeva ~E.\,V., Zhukov~M.\,Yu., Zhukova~N.\,M.
Mathematical Model of a pH-gradient Creation at~Isoelectrofocusing. Part III. Numerical Solution of the Non-stationary Problem.
arXiv:

\bibitem{Part4}
Shiryaeva ~E.\,V., Zhukov~M.\,Yu., Zhukova~N.\,M.
Mathematical Model of a pH-gradient Creation at~Isoelectrofocusing. Part IV. Numerical Solution of the Non-stationary Problem.
arXiv:



\bibitem{SakhVladZhuk}
Sakharova~L.~V., Vladimirov~V.~A., Zhukov~M.~Yu.  Anomalous pH-gradient in Ampholyte Solution. arXiv: 0902.3758vl [physics.chem-ph]. 2009.

\bibitem{SakhSKNC}
Sakharova~L.\,V. Investigation of transformation Gaussian distribution of the concentration at anomalus regimes of isoelectrofocusing. Izvestiya Vyshih Uchebnih Zavedenii. Severo-Kavkazskii Region. Estestvennye Nauki, 2012. Rostov-on-Don. 2012. P. 30--36.

\bibitem{SakhOrel}
Sakharova~L.\,V. Solution of stiff integral-differential IEF problem with help tangent method. Scientific Notes of Orel State University. 2012, No. 6(50). Orel. P. 48--55.

\bibitem{Zhukov2005}
Zhukov~M.\,Yu. Masstransport by an electric field. RGU Press, Rostov-on-Don. 2005.


\bibitem{Fedoruk}
\textsl{Fedoruk~M.\,V.} Asymptotic: Integrals and Series. Ìoscow.: Nauka, 1987.




\end{thebibliography}
\end{document}